\documentclass[reprint,twocolumn, superscriptaddress,nofootinbib,amsmath,amssymb,aps,floatfix]{revtex4}
\usepackage{graphicx} 
\usepackage{subcaption} 
\usepackage{ragged2e} 
\usepackage{enumitem}

\usepackage{bm}

\begin{document}


\title{Local environmental dependence on weak-lensing shear statistics}

\author{Sonia Akter Ema}
\email{sonia.ema@ewubd.edu}
\affiliation{ Department of Mathematical and Physical Sciences, East West University, Aftabnagar, Dhaka 1212, Bangladesh} 
\affiliation{Sydney Institute for Astronomy, School of Physics, A28, The University of Sydney, NSW 2006, Australia}

\author{Md Rasel Hossen}
\email{rasel.hossen@juniv.edu}
\affiliation{Sydney Institute for Astronomy, School of Physics, A28, The University of Sydney, NSW 2006, Australia}
\affiliation{Department of Physics, Jahangirnagar University, Savar, Dhaka 1342, Bangladesh} 

\author{Krzysztof Bolejko}
\email{krzysztof.bolejko@utas.edu.au}
\affiliation{School of Natural Sciences, College of Sciences and Engineering, University of Tasmania, Private Bag 37, Hobart TAS 7001, Australia
}

\author{Geraint F. Lewis}
\email{geraint.lewis@sydney.edu.au}
\affiliation{Sydney Institute for Astronomy, School of Physics, A28, The University of Sydney, NSW 2006, Australia}


\date{\today}

\begin{abstract}
Despite the assumption that an ideal FLRW observer is not dependent on the local environment, observations are biased by the positions of the observers due to the matter correlations in the large-scale structure (LSS) of the universe. The variation of the mass distribution of the LSS of the universe implies that observers residing in different locations may suffer bias in their measurements when they look at the images of distant galaxies. Here, we assess the influence of the local environment on weak gravitational lensing (WL) shear statistics in the context of relativistic $N$-body code, \texttt{gevolution}. We derive numerical constraints on the cosmological parameters from the WL shear angular power spectrum and comment on the local environment's influence on WL shear. We find tighter constraints on the parameter $\Omega_\mathrm{m}$ above redshift $z$ = 0.2, which implies over this redshift the local environment's impact is minor. We also investigate the bispectrum and conclude that on average the impact of the local environment on $f_{\rm NL}$ (a measure of non-Gaussianities) is minimal and consistent with zero effect. However, we find that within the assembly of all possible observers/locations, there will also be a few that could infer the parameter $f_{\rm NL}$ of the order 10. These results could thus be used to estimate the uncertainty in the inference of cosmological parameters such as $f_{\rm NL}$ based on WL shear bispectrum and thus may have implications for future surveys requiring precision at the percent level.
\end{abstract}

\maketitle

\section{Introduction}
\label{sec:intro}
Weak gravitational lensing (WL) refers to the slight distortion of the light coming from distant sources caused by the matter inhomogeneity of the large-scale structure (LSS) of the universe between the source and the observer; for reviews, please see \citep{2015RPPh...78h6901K, 2018ARA&A..56..393M, 2020A&ARv..28....7U, 2010RSPTA.368..967E, 2004bdmh.confE.107S}. It is believed that WL and LSS surveys \citep{2016arXiv161100036D, 2017MNRAS.465.1454H, 2016PhRvD..94b2001A, 2012arXiv1211.0310L, 2011PASA...28..215N, 2018LRR....21....2A} can be powerful tools to extract better constraints on fundamental cosmological parameters. WL allows to study of both nonlinear and non-Gaussian perturbations on a small angular scale \citep{2009MNRAS.395.2065T} and there has been an enormous advancement in this field in the last few decades.


\par Modern cosmology relies heavily on large-scale cosmological $N$-body simulations \citep{2022LRCA....8....1A}. These simulations are the driving force behind most of our theoretical understanding of the formation of cosmic structures, such as galaxies and their distribution, from the early universe to the present day \citep{2022LRCA....8....1A, 2012PDU.....1...50K}. Assuming an extended redshift distribution for the source galaxies, \cite{2019JCAP...10..011W} analysed the convergence maps from the outputs of $N$-body simulations and investigated the impact of baryons on both the current and future WL surveys. By ignoring the influence of baryons and neutrinos, \cite{2015MNRAS.450.2857H} studied the effect of simulation volume on the two-point and four-point WL statistics. 
The vast majority of large-scale cosmological $N$-body simulations are conducted under Newtonian gravity-based assumptions to study the evolution of WL statistics. However, relativistic simulations are important for understanding WL statistics because they give the most accurate and complete picture of how gravity works on cosmological scales, covering non-linear effects and baryonic physics. These models can be used to make accurate predictions that can then be compared with real data to see how well gravity theories and cosmology work. In the presence of relativistic sources, Einstein-based gravity may be required to describe the complex process of nonlinear structure formation in cosmology. \texttt{gevolution} \citep{2016NatPh..12..346A, 2016JCAP...07..053A} is the first $N$-body code for cosmology based on Einstein's theory of general relativity. Based on \texttt{gevolution}, \cite{2019JCAP...12..011H} presented another code, \texttt{k-evolution}, where clustering dark energy was included, \cite{2019JCAP...07..035R} introduced \texttt{fRevolution} that can be used to study the non-linear structure formations in \textit{f(R)} gravity, \cite{2020MNRAS.497.2078L} analysed WL properties by solving Sachs optical equations, and \cite{2023arXiv230207857C} developed the very first relativistic code to solve the dynamic symmetron. Recently, \cite{2023arXiv230111854Q} combined the Particle-Mesh (PM) relativistic \texttt{gevolution} code with the \texttt{TreePM Gadget-4} code to examine the effects of general relativity in cosmology. The main \texttt{gevolution} code was developed by \cite{2016NatPh..12..346A, 2016JCAP...07..053A} and this study uses that code to generate the weak gravitational potentials. 

\par Cosmic shear, an observable consequence of WL, has been identified as a competitive cosmological probe since its first detection \citep{2000MNRAS.318..625B, 2000astro.ph..3338K, 2000A&A...358...30V, 2000Natur.405..143W}. The analysis of cosmic shear has become a key concern in studying the LSS of the universe due to its sensitivity to cosmological parameters. In reference \citep{2011MNRAS.411.2241M}, they developed analytical methods to investigate higher-order WL statistics (for example, shear and flexion) directly through spinorial objects and demonstrated that these higher-order statistical descriptors are effective tools for the investigation of non-Gaussianity in the higher dimensional gravity \citep{2004astro.ph..9224B} offering key scientific drivers for the next generations of WL surveys. 
By analysing the constraints on cosmological parameters from WL shear data, \cite{2021MNRAS.505.4935H} has recently described why the cosmic shear is sensitive to the combination of cosmological parameters $S_8 \propto \sigma_8 \Omega_\mathrm{m}^{0.5}$ but not to the Hubble tension \textit{$H_0$}.

\par While residing on the LSS, the observers' view can be biased due to the variation of density perturbations in the cosmic structures. Previously, a theoretical analysis was done by \cite{2019MNRAS.486.5061R} to the local environmental dependence on the view of an observer.
A recent study \citep{2023arXiv230204507E} investigated the adjustments to the cosmic shear angular power spectra that need to be accounted for when conducting inference using Euclid-like surveys. They evaluated twenty-four correction effects and determined that the local environment effect, and the effects of magnification bias, reduced shear approximation, source-lens clustering, flat universe assumption, and source obscuration, could be significant for Euclid-like surveys. They also concluded that each effect produces significant ( $>$ 0.25$\sigma$) cosmological biases to infer for constraining the cosmological parameters. Upcoming large-scale photometric studies, like as the Dark Energy Survey (DES) and Euclid, will be crucial for understanding the changes in galaxy characteristics over time and the influence of the local environment. In reference \cite{2015MNRAS.451..660E}, they investigated the impact of the local environment, considering the Sloan Digital Sky Survey (SDSS) spectroscopic and photometric redshift data and also simulation data. Their analysis reveals that photometric settings have a narrower range of variation compared to spectroscopic data. This is because galaxies with uncertain redshifts tend to disperse from dense environments to less dense ones. In our recent work \citep{2022MNRAS.509.3004E}, we studied the influence of the local environment due to the density fluctuations in cosmic structures based on the WL convergence data. We have calculated the probability distribution functions (PDFs) and angular power spectrum from WL convergence map data and thoroughly analysed how the observers' view on the LSS impacted due to the density variation. We also computed the constraints on cosmological parameters from the angular power spectrum of WL convergence there and commented on how the local environment influences the  WL convergence. The current study is distinguished from previous work by its focus on the weak-lensing observable (shear) and on the influence of the local environment on higher-order statistics (bispectrum) and the non-Gaussian parameter.

Here, we numerically examine how the locality of the observer influences the WL shear statistics. We analyse how the WL shear PDF and angular power spectrum, as well as higher-order statistics (i.e., shear bispectrum), depend on the local environment of the universe. We also study the constraints on cosmological parameters and discuss the influence of the local environment on WL shear. We consider the $\Lambda$CDM cosmology as a background where the snapshots for particle's positions and weak gravitational potentials are generated from \texttt{gevolution} \citep{2016JCAP...07..053A, 2016NatPh..12..346A} and then we solve geodesic equations using the \texttt{3D Ray Bundle Tracers (3D-RBT)} algorithm \citep{2022MNRAS.509.3004E, 2022MNRAS.513.5575H, 2022MNRAS.509.5142H}. Finally, we compute the PDFs, angular power spectrum, and bispectrum from WL shear map data and elaborately discuss the local environmental dependence on WL shear. 


This paper is organised as follows: Section \ref{sec:Background and Algorithms} contains brief descriptions of WL theory and our methodology to calculate the WL properties. In Section \ref{sec:Numerical techniques} we mention the details of the $N$-body simulations that we use in this work and how we find cosmic structures from the simulation volume by using publicly available algorithms. Section \ref{sec:WL Shear Angular power spectrum and bispectrum} focuses the discussion on how we infer the WL shear angular power spectrum and bispectrum theoretically as well as numerically. Section \ref{sec:Results} is devoted to describing the results of this study and finally, we conclude in Section \ref{sec:Discussion and outlook}.

\section{Background and Algorithms}
\label{sec:Background and Algorithms}

\subsection{Theoretical background of WL}
Cosmic shear is related to the image distortion of distant galaxies caused by intervening the LSS. The explicit expression of 3D cosmic shear at an arbitrary position can be expressed by the second-order derivatives of the effective lensing potential $\psi_{ij}$ as

\begin{equation}
\psi_{\mathrm{ij}} = \frac{D_\mathrm{d} D_{\mathrm{ds}}}{D_\mathrm{s}} \frac{2}{c^2} \int\frac{\partial^2
\phi(x_3)}{\partial x_\mathrm{i} \partial x_\mathrm{j}}dx_3,
\label{psiij}
\end{equation}
 
\noindent here D$_\mathrm{d}$, D$_\mathrm{{ds}}$, and D$_\mathrm{s}$ are the angular diameter distances from the observer to the lens, the lens to the source, and the observer to the source, respectively, \rm c is the velocity of light, $x_i$ and $x_j$ are position coordinates (here i, j = 1, 2, 3 that represent the directions), and $\phi$ is the gravitational potential. 

For multiple deflections, the shear tensor, ${\cal U}^{(\mathrm{i})}$, can be expressed in terms of the deflector index $i$

\begin{equation}
{\cal U}^{(\mathrm{i})} = \left( \begin{array}{cc}
	\psi_{11}^{(i)}  & \psi_{12}^{(i)} \\ [0.6cm]
	\psi_{21}^{(i)}   & \psi_{22}^{(i)}
	\end{array}
	\right).
\label{U2}
\end{equation}

\noindent The multiple lens-plane method \citep{1992grle.book.....S} is useful to calculate the final properties of WL statistics at $z=0$ from the combination of ${\cal U}^{(\mathrm{i})}$ for all deflectors. The Jacobian matrix describes the change of the shape of the source on the image plane and the final form of Jacobian is 

\begin{equation}
{\cal A} = \left( \begin{array}{cc}
	1-\psi_{11}  & -\psi_{12} \\ [0.3cm]
	-\psi_{21}   & 1-\psi_{22}
	\end{array}
	\right),
\label{calA}
\end{equation}

\noindent in the case of WL, the components of the overall 2D shear,
$\gamma$, are
\begin{equation}
\gamma_1 = \frac{1}{2}(\psi_{11}-\psi_{22}),
\label{gamma1}
\end{equation}
and
\begin{equation}
\gamma_2 = \psi_{21} = \psi_{12}.
\label{gamma2}
\end{equation}

\noindent In a weak shear field, the 2D shear is 
\begin{equation}
\gamma = \gamma_1 + i\gamma_2.
\end{equation}

\noindent The convergence that can be measured from the integrated mass density along the line of sight, can be expressed as 
\begin{equation}
\kappa = \frac{1}{2}(\psi_{11}+\psi_{22}),
\label{kappa1}
\end{equation}

\noindent and the magnification is 
\begin{equation}
\mu =\left|\det \cal A \right|^{-1} = \frac{1}{|(1-\kappa)^2-\gamma^2|}.
\label{mu1}
\end{equation}

\noindent Shear is obtained from the stretched axes of an elliptical image, where
\begin{equation}
a = (1-\kappa - \mid \gamma \mid)^{-1}~~{\rm (semi-major~~axis),}
\end{equation}
and
\begin{equation}
b = (1-\kappa + \mid \gamma \mid)^{-1}~~{\rm (semi-minor~~axis)}.
\end{equation}

\subsection{3D ray bundle tracers (3D-RBT)}

The ray-shooting method (RSM) is a well-known method for measuring the lensing statistical quantities \citep{1986ApJ...301..503P, 1986A&A...166...36K}. In this study, we use another method, namely, the ray bundle method (RBM) to measure the WL statistics numerically. The original RBM that was developed by \cite{1999MNRAS.306..567F} follows a general format and it's easy to include a variety of cosmological lens models in this method \citep{2002MNRAS.331..180F, 2011MNRAS.416.1616F, 2012MNRAS.420..155K}. RBM computes cosmic shear directly from the deformed shape of the ray bundles, which is not possible with RSM. The main structural difference between RSM and RBM is: RSM models single light ray in individual line of sight whereas, RBM models a bundle of light rays (that consists of eight light rays and a central ray) in each line of sight. The main advantage of using RBM is - we can easily compute the correspondence between the source and the image from the initial and final shapes of each ray bundle. In this study, we use a ray tracing algorithm, $\texttt{3D-RBT}$ by adopting RBM and analysing WL statistics. Here we give below a brief overview of the algorithm $\texttt{3D-RBT}$, for more details we refer the readers to our recently published works \citep{2022MNRAS.509.3004E, 2022MNRAS.509.5142H, 2022MNRAS.513.5575H}.

\begin{itemize}
    \item \texttt{generating snapshots:} first, we run the relativistic $N$-body code \texttt{gevolution}\footnote{{\protect\url{https://github.com/gevolution-code/gevolution-1.2}}} \citep{2016JCAP...07..053A, 2016NatPh..12..346A} and then save snapshots of weak gravitational potentials at different redshifts. 
    
    \item \texttt{identifying cosmic structures:} we identify haloes and voids from the particle's snapshot (that is generated and then saved by running the code \texttt{gevolution}) at today's redshift. We adopt the public code \texttt{ROCKSTAR}\footnote{\protect\url{https://github.com/yt-project/rockstar}} to identify the positions of haloes and another public code \texttt{Pylians}\footnote{\protect\url{https://github.com/franciscovillaescusa/Pylians}} to identify the positions of voids. 
    
    \item \texttt{setting observers' position and initial conditions:} 
    once the locations of cosmic structures have been determined, we proceed to classify all haloes based on their masses and voids based on their radii. Next, we locate the observers at the centre of the distinct haloes and voids and set up discrete initial instances for each observer.
    
    \item \texttt{projecting ray bundles, calculating Christoffel symbols,  and solving null geodesics:} we project light rays as bundles in different directions on the \texttt{healpy}\footnote{{\protect\url{https://github.com/healpy/healpy}}} sphere. Here each bundle consists of in total nine light rays/geodesics, where one centred geodesic is surrounded by eight geodesics. We then calculate light paths for each observer through the calculation of Christoffel symbols and then solve the null geodesic equations as bundles \citep{2011MNRAS.412.1937B, 2011JCAP...02..025B}.
    
    \item \texttt{fitting ellipses and measuring WL properties:} due to the LSS of the universe, the initial circular shape of the ray bundles get distorted and we fit the data of distorted images into ellipses using an ellipse fitting algorithm. We run more than a hundred ray tracing simulations and then calculate the WL properties. The ways of measuring WL properties are:
    
    \begin{itemize}
        \item \texttt{i) magnification:} magnification comes from the bending of light emitted by distance sources. Using this $\texttt{3D-RBT}$ method, we calculate magnification from the initial and final shapes of the ray bundles. The formula is 
        \begin{equation}
        \mu = \frac{A_{\mathrm{image}}}{A_{\mathrm{source}}},
        \label{mu_RBM}
        \end{equation}
        \noindent where $A_{\mathrm{image}}$ and $A_{\mathrm{source}}$ are the area of the image and the area of the source respectively. For a perturbed universe, $\mu < 1 $ indicates demagnification and that occurs in under-dense regions whereas, $\mu > 1 $ refers to magnification and that happens in over-dense regions.
        
        \item \texttt{ii) cosmic shear:} it is possible to measure cosmic shear for each bundle from the image distortion of each source using the $\texttt{3D-RBT}$ method. For each ray bundle, we obtain WL shear from the alignment of an ellipse. We use the following equation to calculate the WL shear here
        
    \begin{equation}
    \gamma = \frac{b - a}{b + a},
    \label{shear_RBM}
    \end{equation}

    \noindent here b is the semi-minor and a is the semi-major axes of the ellipse. 
    \end{itemize}
     
\end{itemize}

\section{Numerical techniques}
\label{sec:Numerical techniques}

\begin{figure}
    \centering
    \includegraphics[width=\columnwidth]{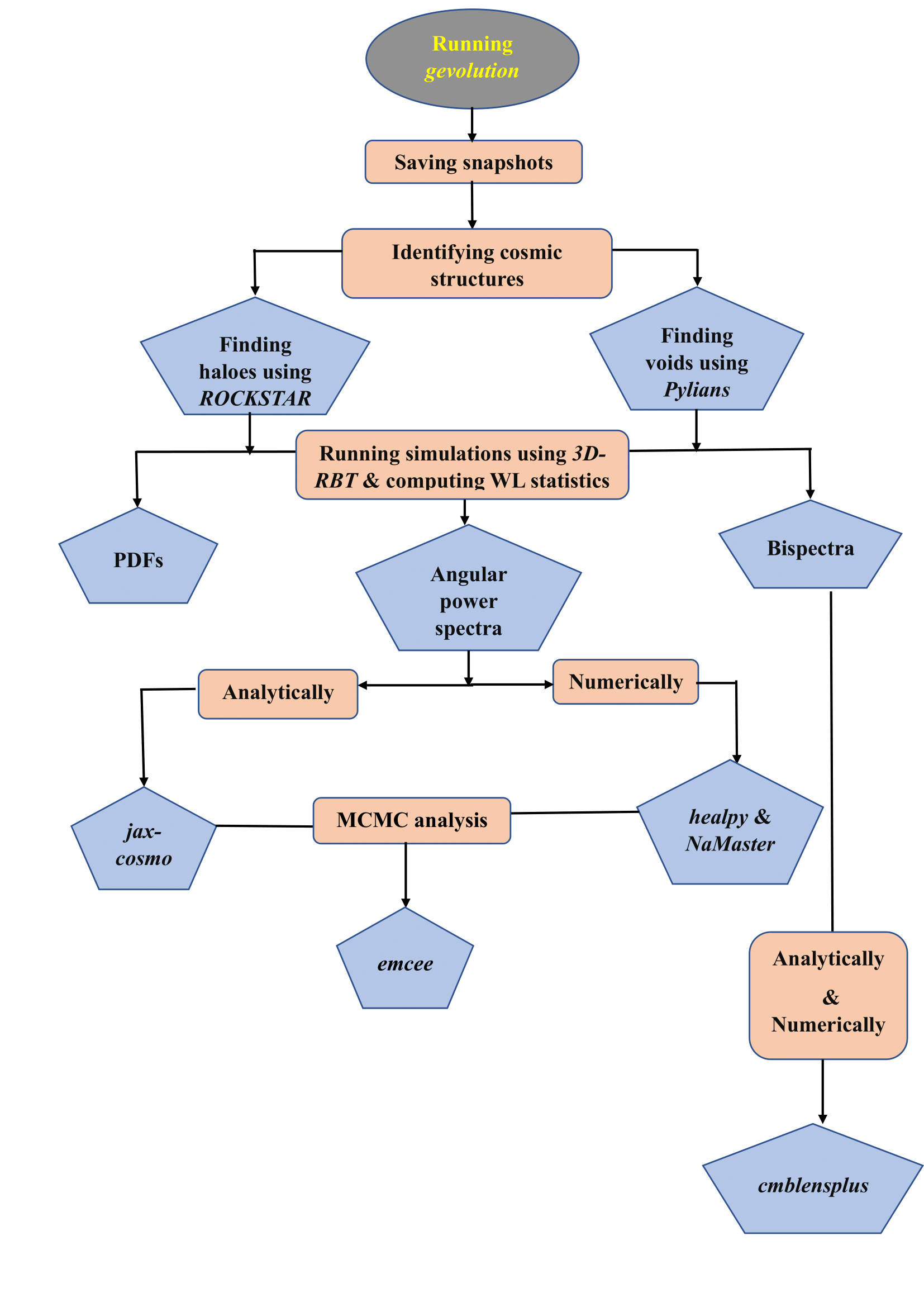}
    \caption{\justifying Schematic diagram summarizing all the packages that we use for different purposes in this paper.}
    \label{figg}
\end{figure}

The pipeline implemented in this paper is summarized in Figure \ref{figg}. Below we discuss various elements of software that comprise our numerical pipeline.

\subsection{$N$-body simulations}
The weak gravitational potentials are generated using the weak field metric derived from the relativistic $N$-body code \texttt{gevolution} \citep{2016NatPh..12..346A, 2016JCAP...07..053A} mechanism. We adopt a $\Lambda$CDM cosmology with the following values of cosmological parameters: $\Omega_\mathrm{m}$ = 0.312, $\Omega_{\Lambda}$ = 0.6879, dimensionless Hubble parameter $h$ = 0.67556, spectral index $n_\mathrm{s}$ = 0.9619, and massless neutrinos with $N_\mathrm{eff}$ = 3.046, respectively. The primordial amplitude of scalar perturbations is fixed at $A_\mathrm{s}$ = 2.215 $\times$ $10^{-9}$ when the pivot scale is $k_*$ = 0.05 $\mathrm{Mpc^{-1}}$, and the parameter $S_8$ is hard-coded in \texttt{gevolution}. The $N$-body simulations used in this analysis start from Gaussian initial conditions, i.e., the primordial density fluctuations follow a Gaussian distribution. The initial redshift of \texttt{gevolution} was $z = 100$ and we ran more than 300 ray tracing simulations using our algorithm \texttt{3D-RBT}, where the number of dark matter particles in each simulation is 256$^3$ having a simulation volume of $(320~\mathrm{Mpc}/h)^3$. The observers are set at today's redshift, and we project the ray bundles up to a comoving distance of $1.5~\mathrm{Gpc}/h$ (that corresponds to redshift $z = 0.62$) in the backward direction of time. We acknowledge that the $N$-body and ray-tracing simulation setups are the same as in Ref. \citep{2022MNRAS.509.3004E}. However, in the present work, we increased the number of realizations in order to achieve improved statistical confidence. This required running additional $N$-body simulations and corresponding ray-tracing analyses compared to Ref.  \citep{2022MNRAS.509.3004E}. To generate the PDFs, angular power spectra, and bispectra for a halo mass range, we run 60 ray tracing simulations and solve 49,152 bundles of null geodesics for different initial conditions by setting observers within 60 different haloes at that specific mass range. Finally, we compute the mean PDFs, angular power spectra, and bispectra and generate the error bars. We repeat this procedure for all halo mass and void radius ranges. We carefully pick the total number of bundles as a multiple of 9 because each bundle consists of 9 light rays. There is no specific reason for taking the number of lines of sight that we chose here, and one can easily run our simulation with different numbers of bundles.

\subsection{Finding cosmic structures}
In the observable universe, haloes and voids are two cosmic structures that are formed due to the non-identical mass distributions in the LSS. We find the positions of haloes and voids from the particle's snapshot of \texttt{gevolution} and then set the observers within these cosmic structures. The following is a brief overview of identifying the cosmic structures.
\begin{itemize}
    \item \texttt{Haloes:} these are higher dense regions in the LSS and we adopt the public code \texttt{ROCKSTAR}\footnote{\url{https://github.com/yt-project/rockstar}} \citep{2013ApJ...762..109B} to find out the positions of haloes in the simulation volume. We obtain a total of 30, 321 haloes from the particle's snapshot of \texttt{gevolution} at today's redshift. All the haloes here don't possess same masses and we find a total of 17,613 haloes whose mass is smaller than $10^{12.5}$ $\mathrm{M}_\odot$ $h^{-1}$, a total of 11,353 haloes when $10^{12.5}$ $\mathrm{M}_\odot$ $h^{-1}$ \(<\) halo mass \(<\) $10^{13.5}$ $\mathrm{M}_\odot$ $h^{-1}$, and in total 1,300 number of haloes where $10^{13.5}$ $\mathrm{M}_\odot$ $h^{-1}$ \(<\) halo mass \(<\) $10^{14.5}$ $\mathrm{M}_\odot$ $h^{-1}$. 
    
    \item \texttt{Voids:} these are considered as the under-dense regions in the LSS of the universe and they are surrounded by galaxy clusters. We use a public code \texttt{Pylians}\footnote{\url{https://github.com/franciscovillaescusa/Pylians}} in our study to identify the positions of voids within the simulation volume. A total of 5,261 voids of different radii are found from the particle's snapshot at $z=0$. Among them, 2,559 voids have a radius of 10 - 20 $\mathrm{Mpc}/h$, 342 voids in the radius range of 21 - 30 $\mathrm{Mpc}/h$, and 126 voids in the radius range of 31 - 40 $\mathrm{Mpc}/h$ have been found. 
    
\end{itemize}

\begin{figure}
     \centering
     \begin{subfigure}[b]{0.42\textwidth}
         \centering
         \includegraphics[width=\textwidth]{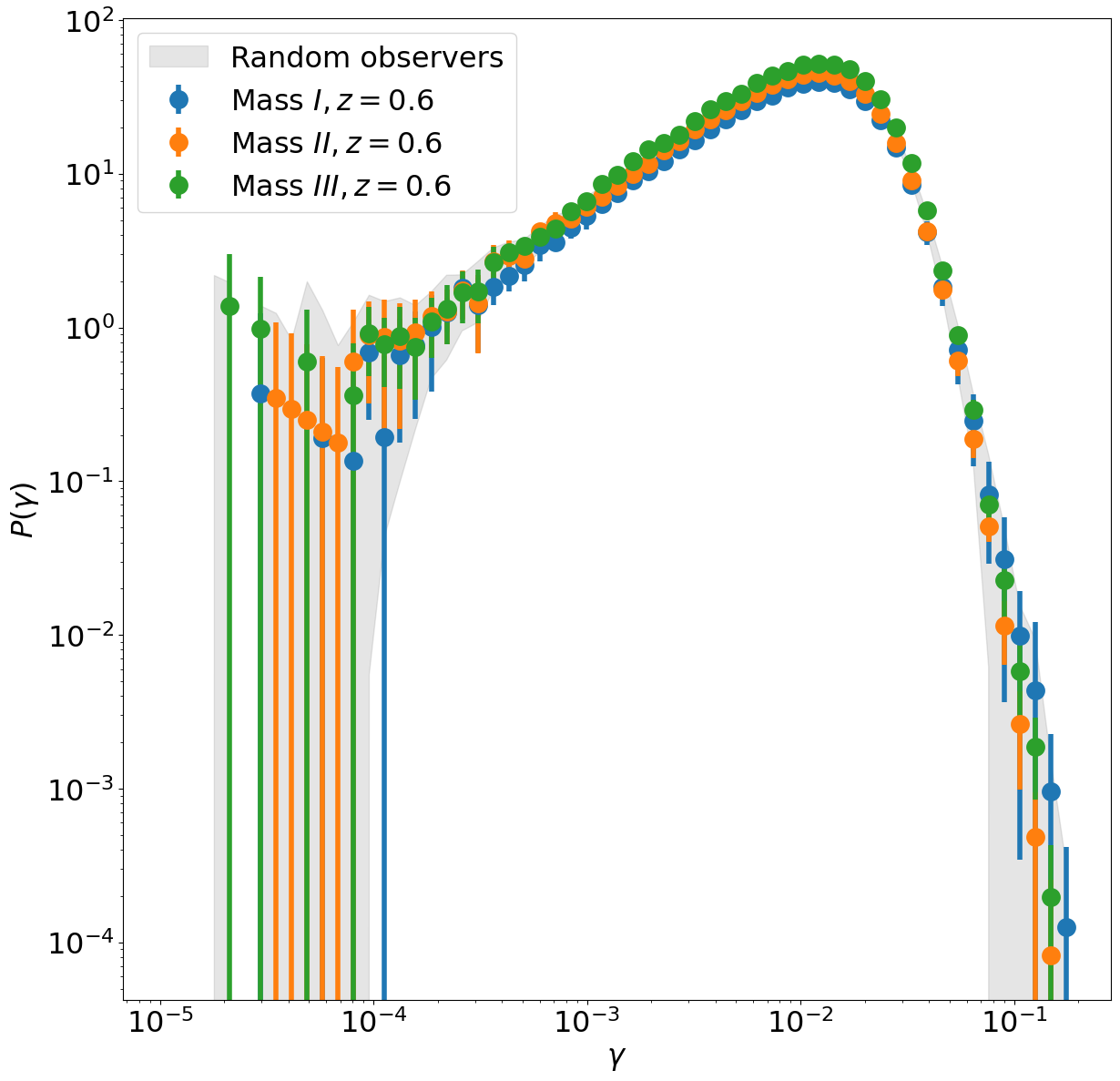}
     \end{subfigure}
     \hfill
     \begin{subfigure}[b]{0.42\textwidth}
         \centering
         \includegraphics[width=\textwidth]{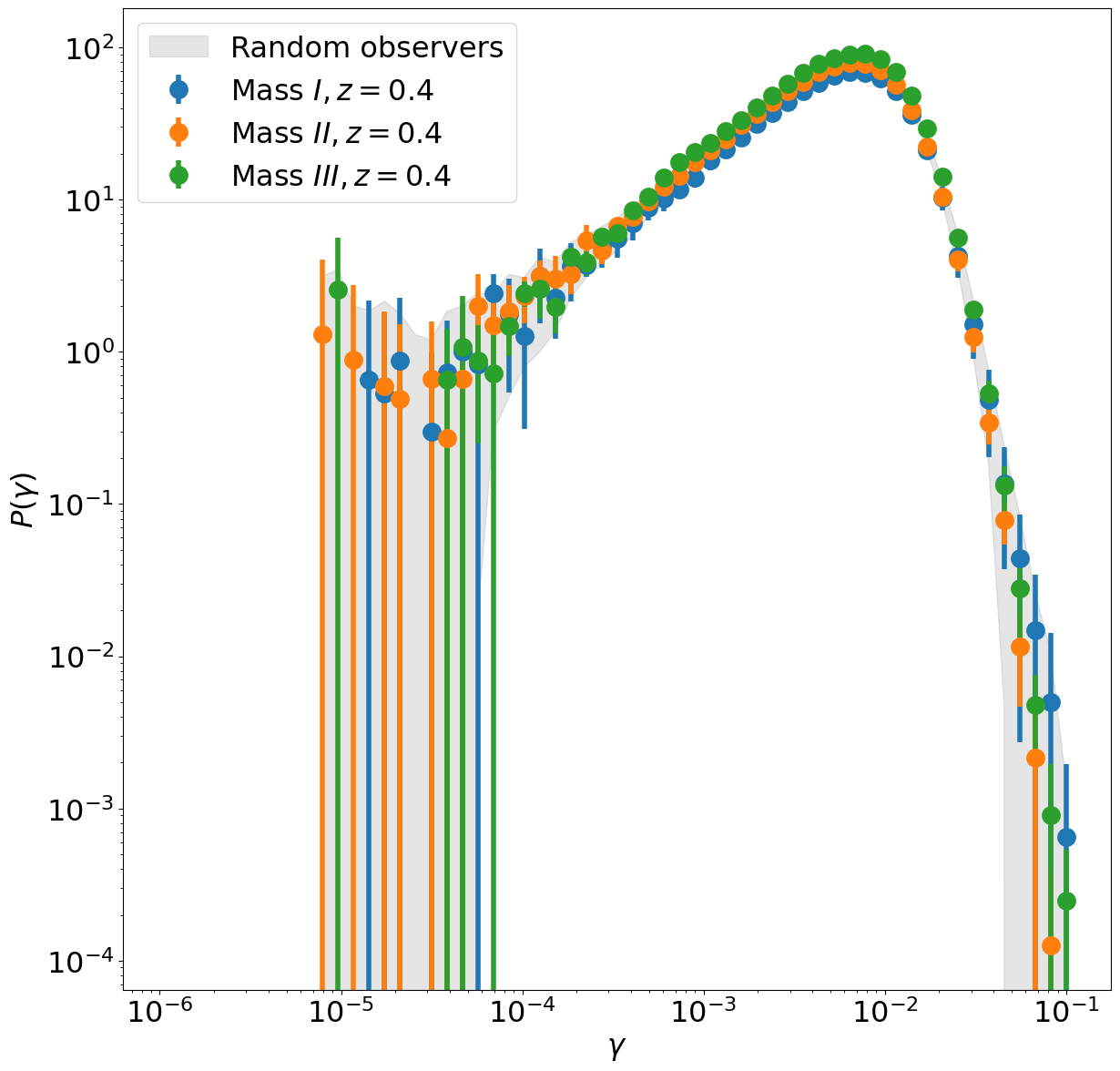}
     \end{subfigure}
     \hfill
     \begin{subfigure}[b]{0.42\textwidth}
         \centering
         \includegraphics[width=\textwidth]{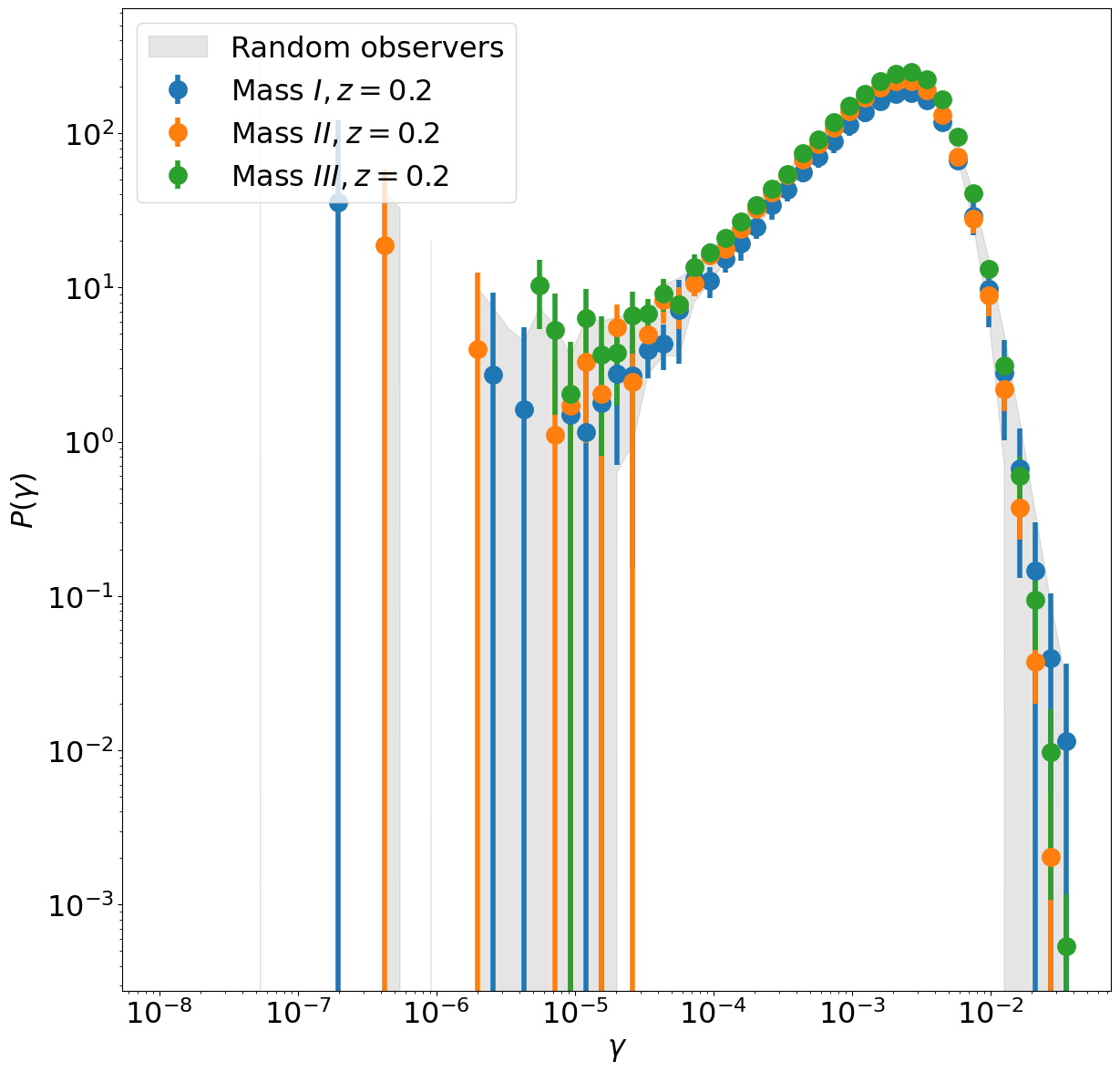}
     \end{subfigure}    
        \caption{\justifying WL shear PDFs with respect to redshift when observers are positioned in haloes with varying masses (Mass I: halo mass \(<\) $10^{12.5}$ $\mathrm{M}_\odot$ $h^{-1}$; Mass II: $10^{12.5}$ $\mathrm{M}_\odot$ $h^{-1}$ \(<\) halo mass \(<\) $10^{13.5}$ $\mathrm{M}_\odot$ $h^{-1}$; Mass III: $10^{13.5}$ $\mathrm{M}_\odot$ $h^{-1}$ \(<\) halo mass \(<\) $10^{14.5}$ $\mathrm{M}_\odot$ $h^{-1}$). Markers in this context represent mean PDFs, whereas error bars of varying colours display data indicating 68\% deviation from the mean shear value within each mass range. The gray-shaded regions indicate the deviation band if the observer is randomly located within the simulation box.}
        \label{fig1}
\end{figure}




\begin{figure}
     \centering
     \begin{subfigure}[b]{0.42\textwidth}
         \centering
         \includegraphics[width=\textwidth]{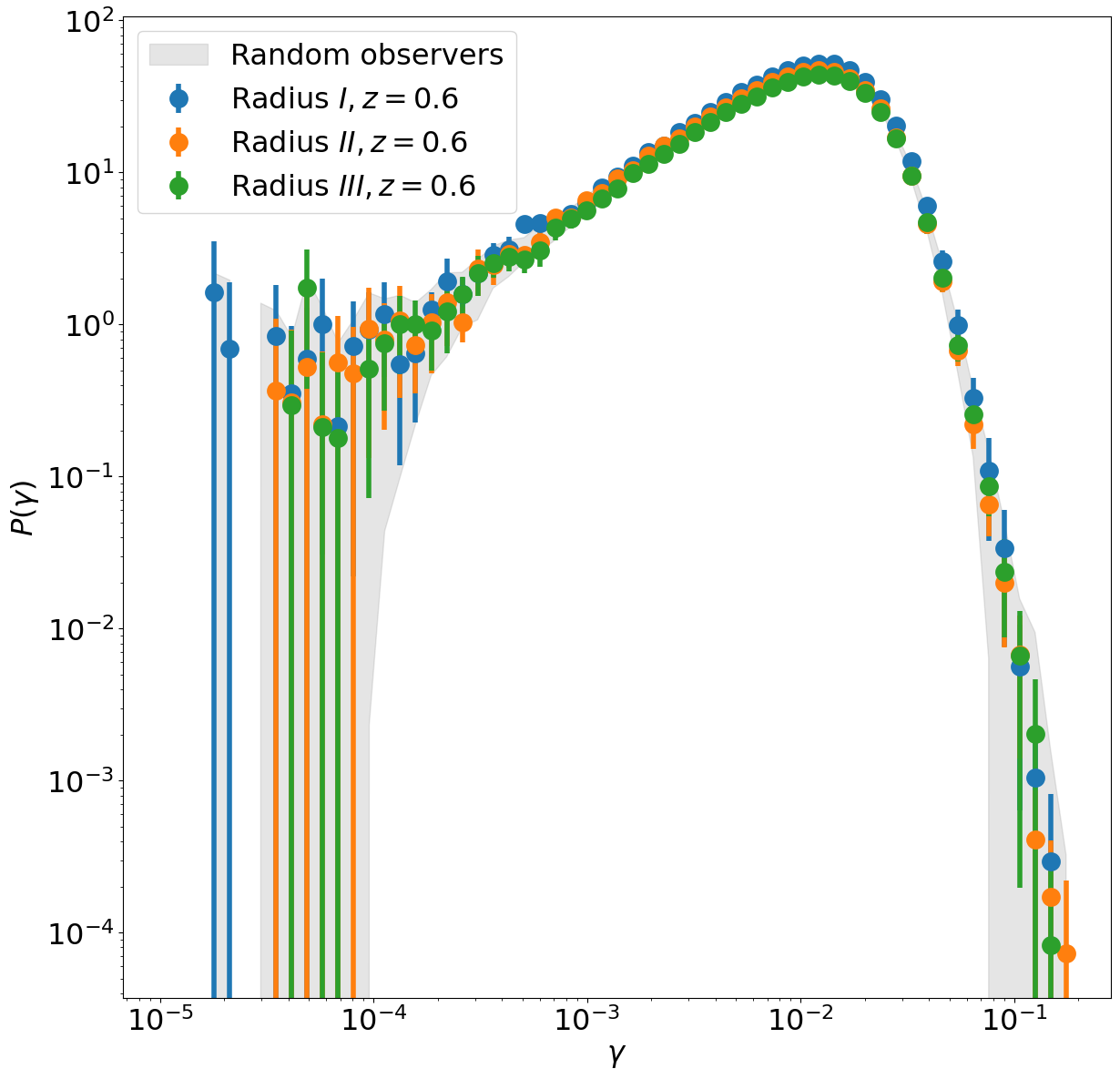}
     \end{subfigure}
     \hfill
     \begin{subfigure}[b]{0.42\textwidth}
         \centering
         \includegraphics[width=\textwidth]{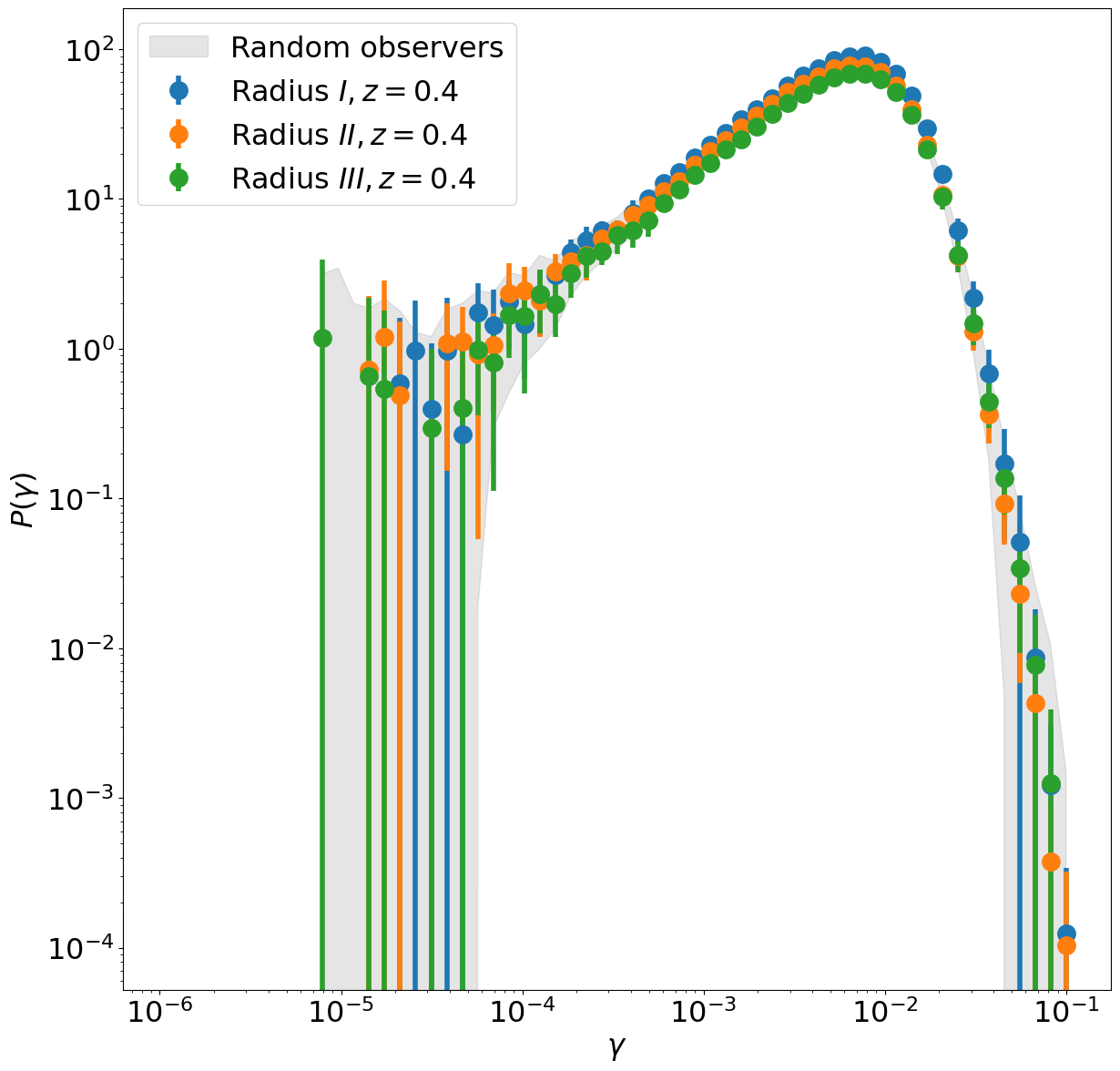}
     \end{subfigure}
     \hfill
     \begin{subfigure}[b]{0.42\textwidth}
         \centering
         \includegraphics[width=\textwidth]{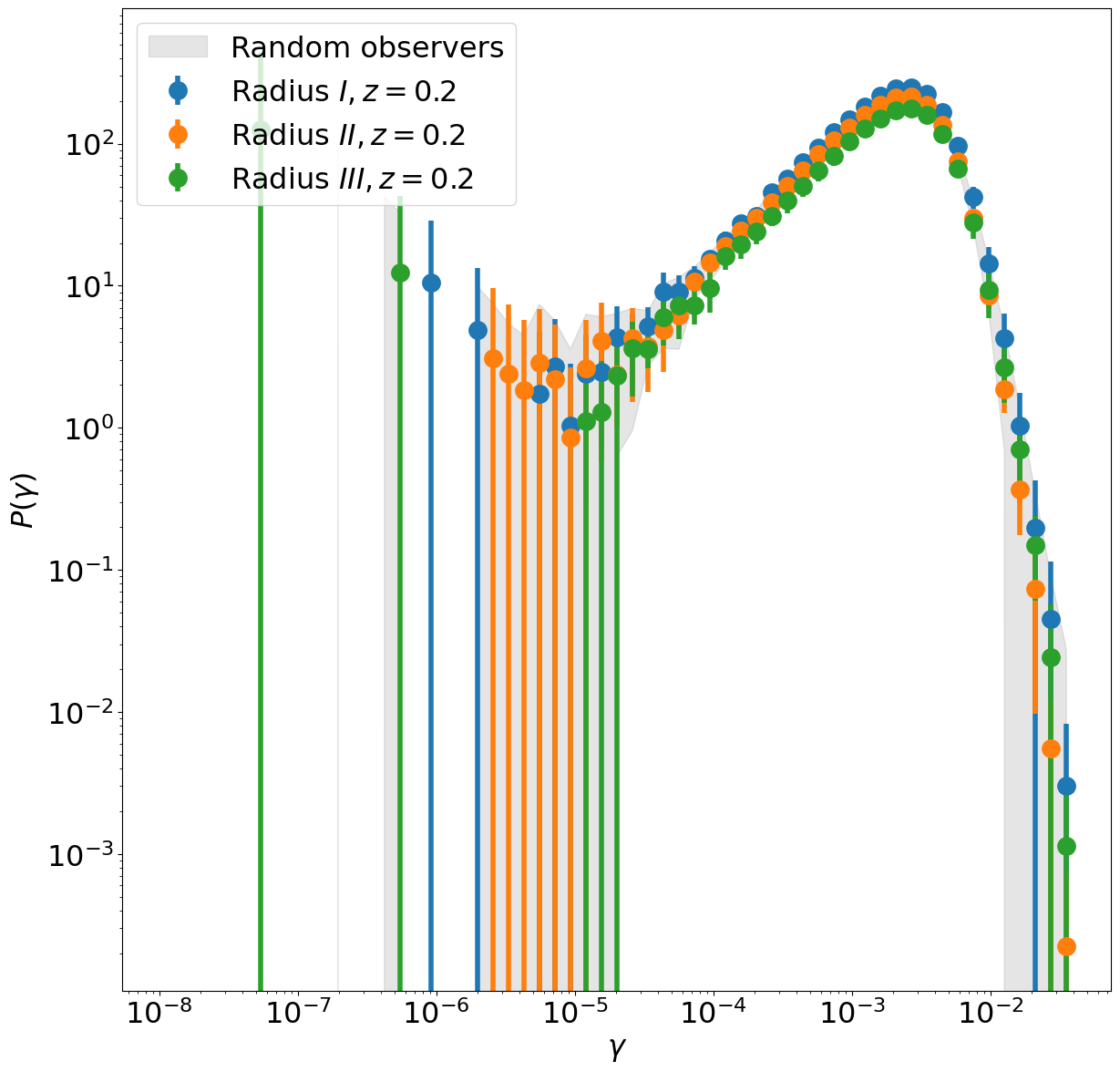}
     \end{subfigure}
        \caption{\justifying Changes in WL shear PDFs as based on redshift when observers were positioned in voids having different radii (Radius I: 10-20 \rm{Mpc}/$h$; Radius II: 21-30 \rm{Mpc}/$h$; Radius III: 31-40 Mpc/$h$). Markers in this context represent mean PDFs, whereas error bars of varying colours display data indicating 68\% deviation from the mean shear value within each radius range. The gray-shaded regions indicate the deviation band if the observer is randomly located within the simulation box.}
        \label{fig2}
\end{figure}

\begin{figure}
     \centering
     \begin{subfigure}[b]{0.42\textwidth}
         \centering
         \includegraphics[width=\textwidth]{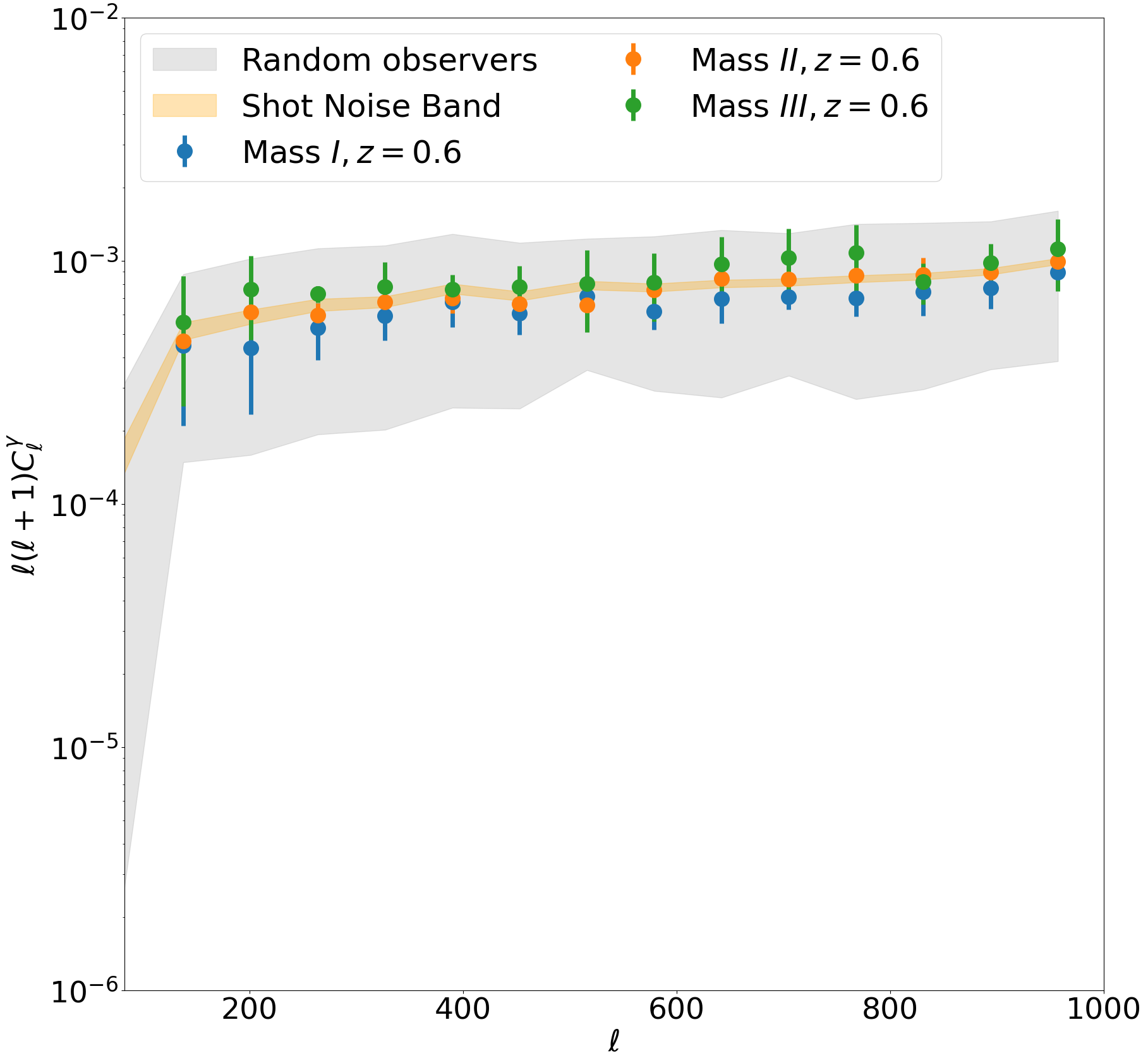}
     \end{subfigure}
     \hfill
     \begin{subfigure}[b]{0.42\textwidth}
         \centering
         \includegraphics[width=\textwidth]{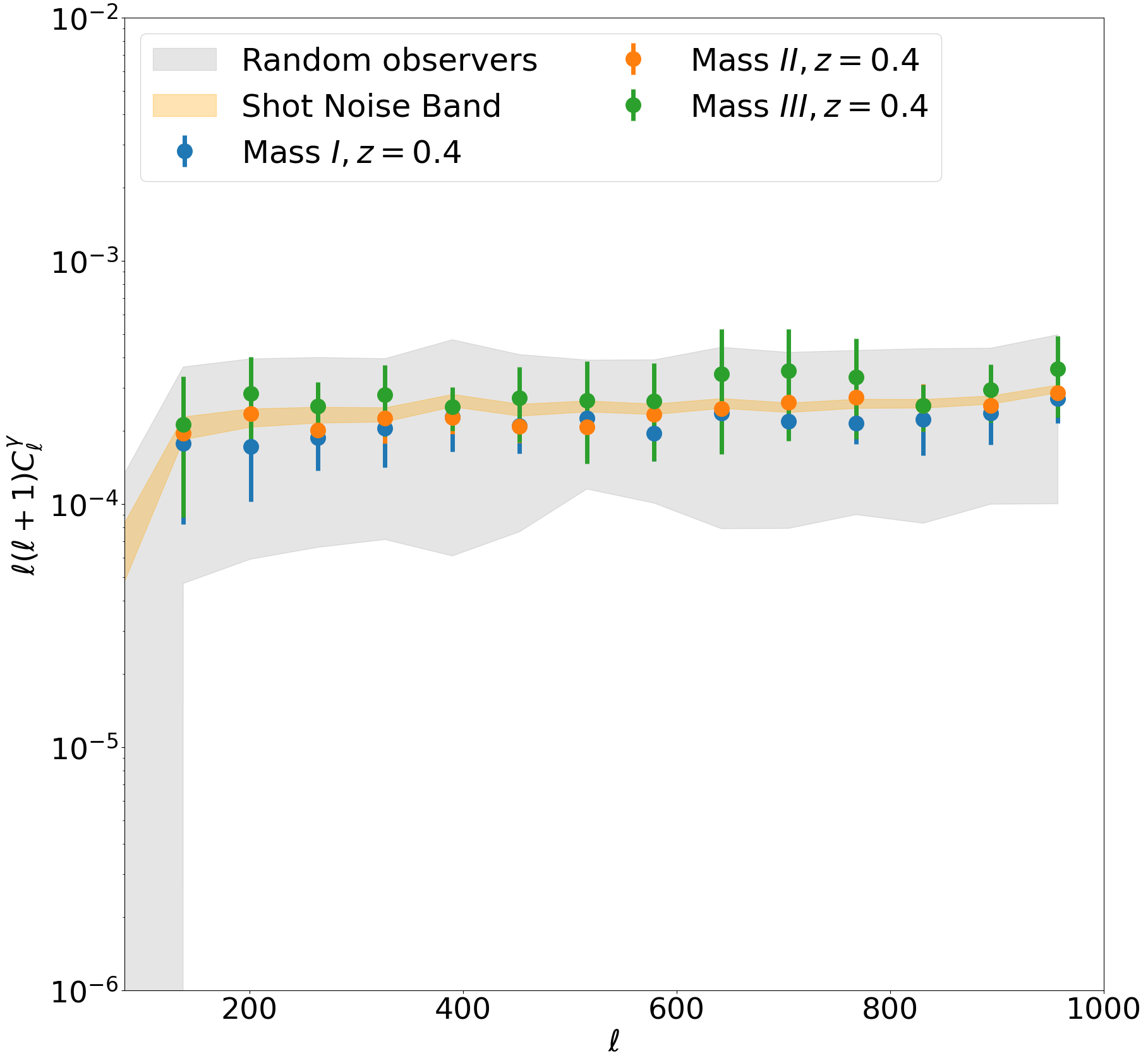}
     \end{subfigure}
     \hfill
     \begin{subfigure}[b]{0.42\textwidth}
         \centering
         \includegraphics[width=\textwidth]{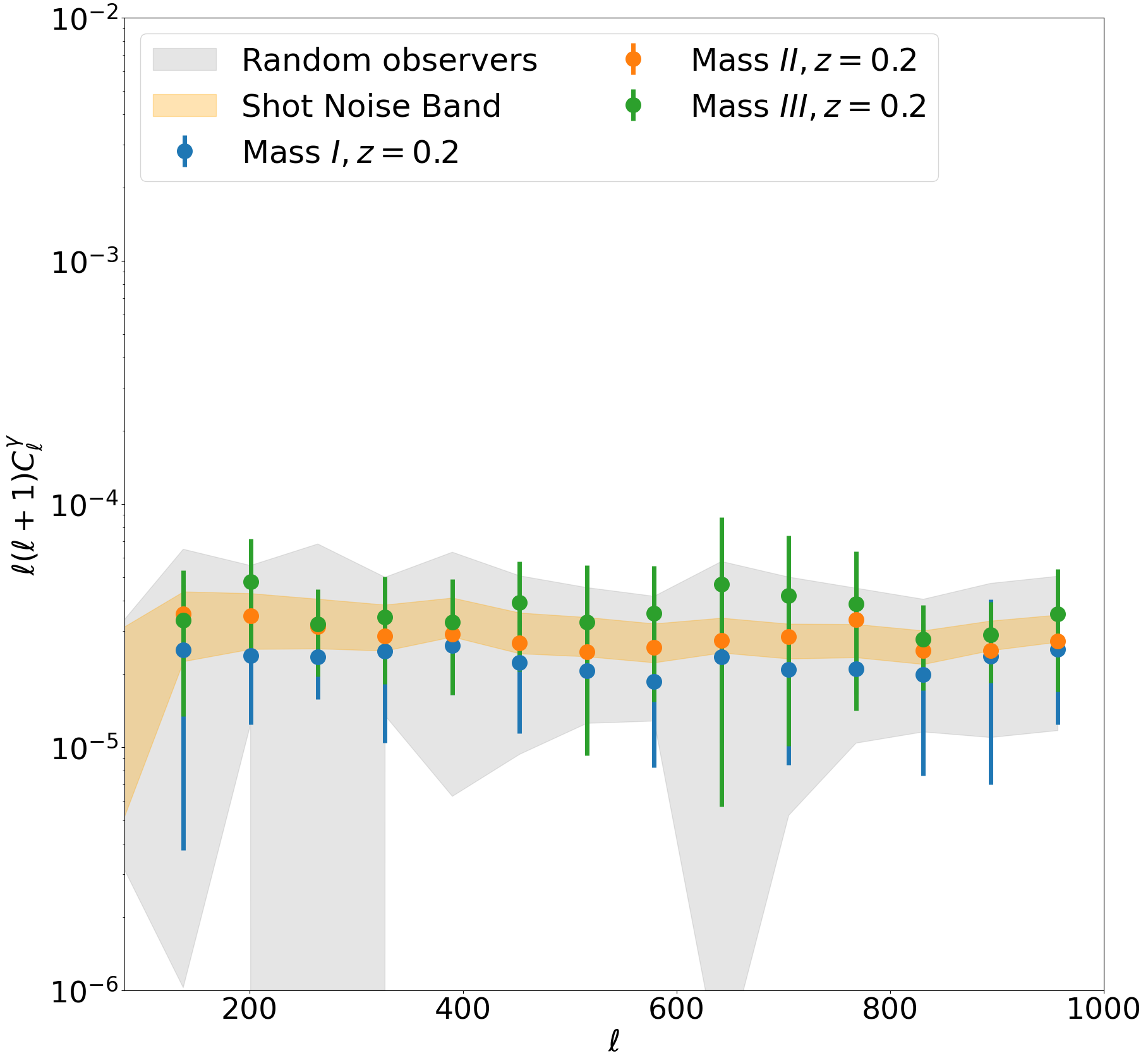}
     \end{subfigure}
        \caption{\justifying Analysis of the shear angular power spectrum with respect to redshift for observers residing in haloes of varying masses. In this context, markers are used to represent mean values, while error bars of varying colours are used to display the data around the mean values for separate redshifts within each mass range. The gray-shaded regions indicate the deviation band if the observer is randomly located within the simulation box. The orange-shaded regions indicate the expected deviation band based on the shot noise contribution to the WL shear angular power spectrum.}
        \label{fig3}
\end{figure}
\begin{figure}
     \centering
     \begin{subfigure}[b]{0.42\textwidth}
         \centering
         \includegraphics[width=\textwidth]{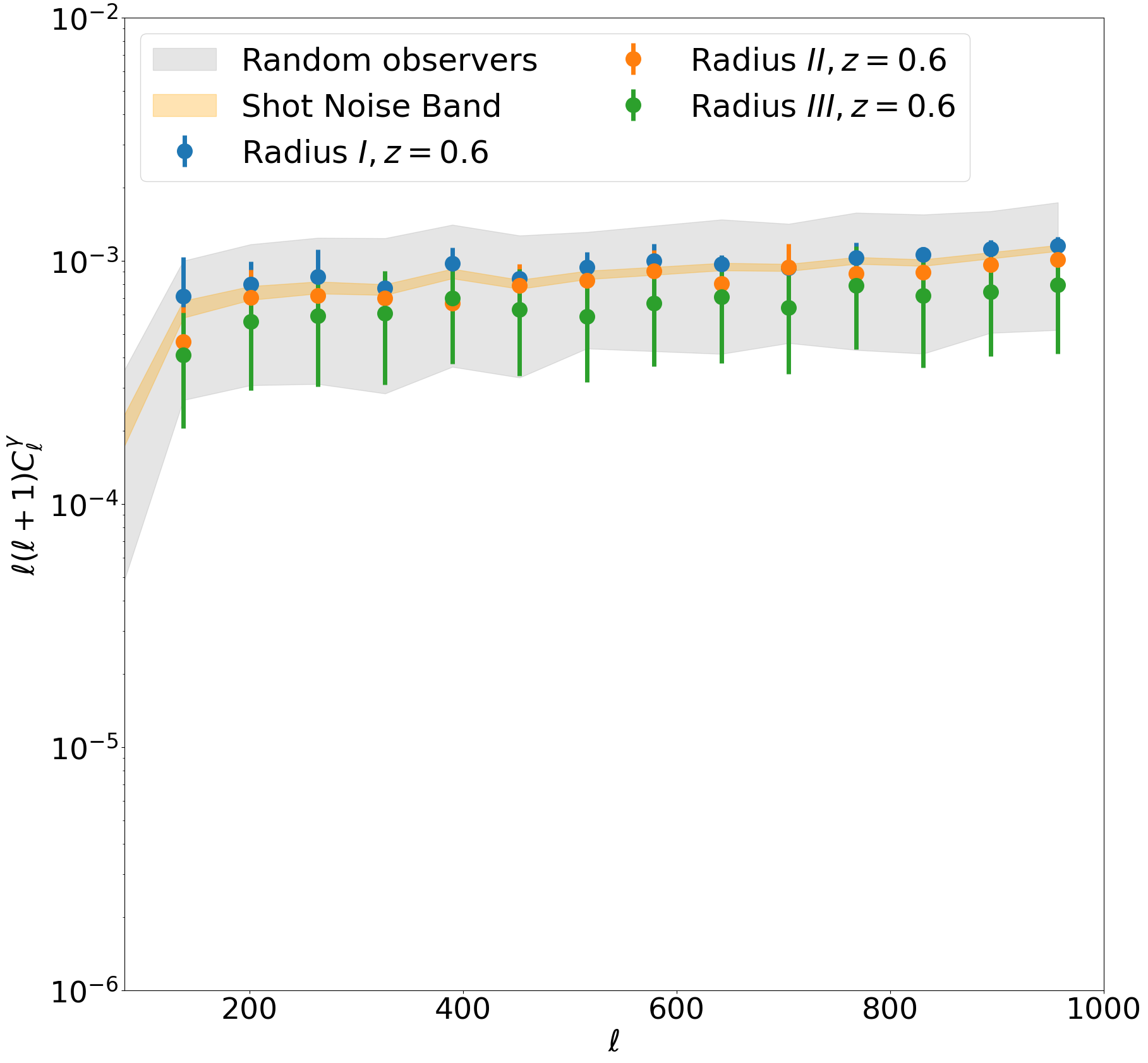}
     \end{subfigure}
     \hfill
     \begin{subfigure}[b]{0.42\textwidth}
         \centering
         \includegraphics[width=\textwidth]{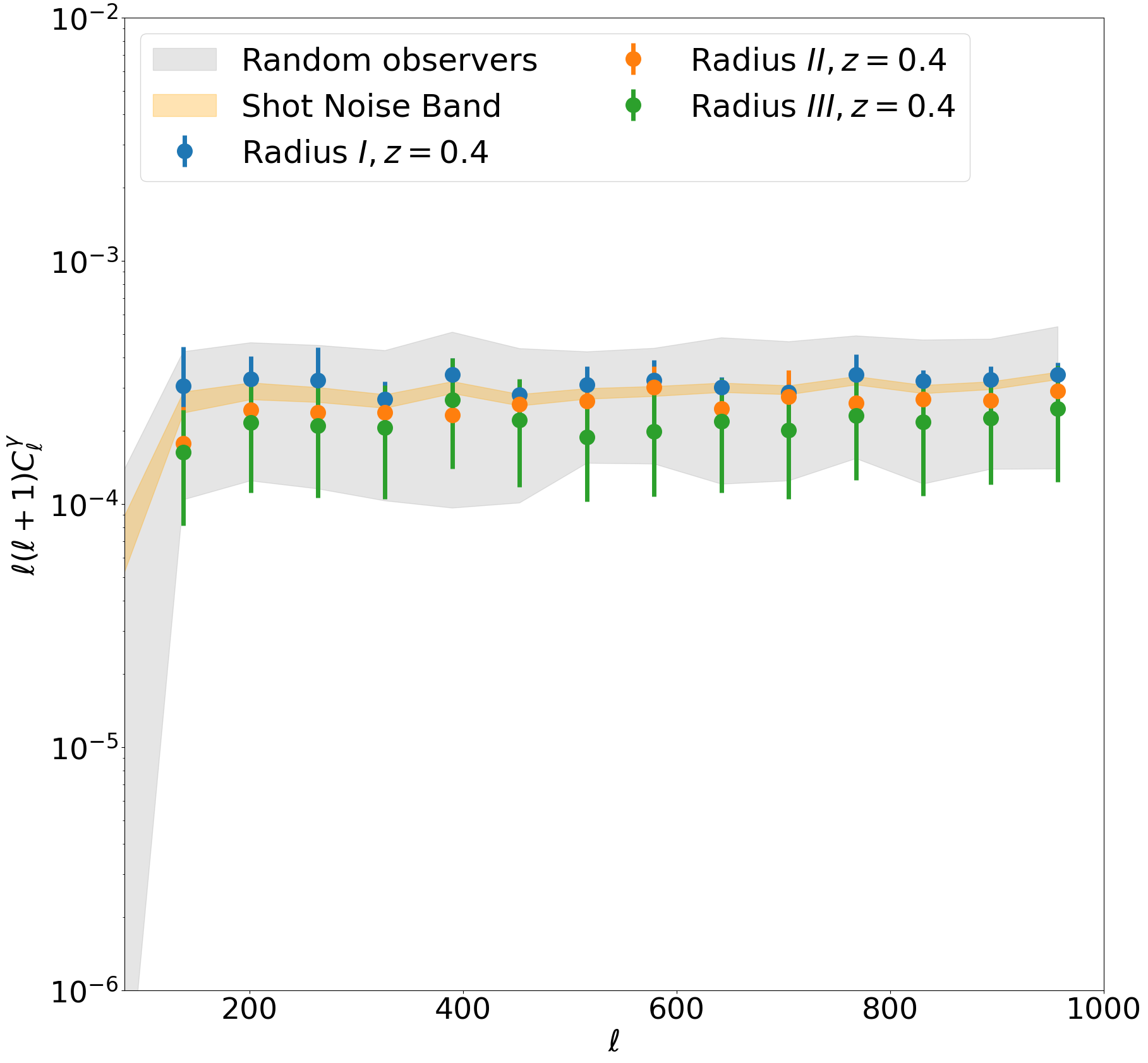}
     \end{subfigure}
     \hfill
     \begin{subfigure}[b]{0.42\textwidth}
         \centering
         \includegraphics[width=\textwidth]{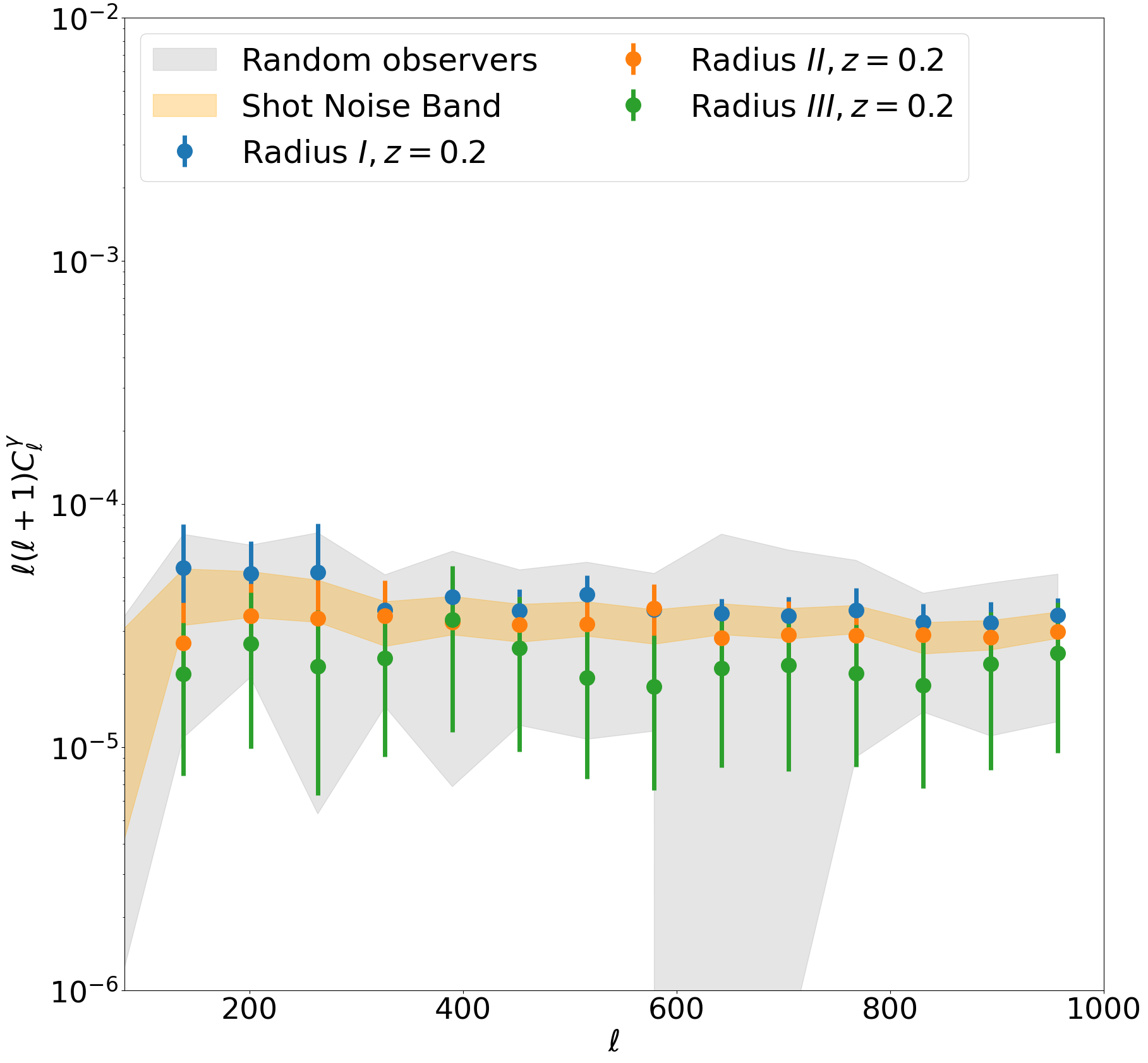}
     \end{subfigure}
        \caption{\justifying Analysis of the shear angular power spectrum with respect to redshift for observers residing in voids of varying radii. In this context, markers are used to represent mean values, while error bars of varying colours are used to display the data around the mean values for separate redshifts within each radius range. The gray-shaded regions indicate the deviation band if the observer is randomly located within the simulation box. The orange-shaded regions indicate the expected deviation band based on the shot noise contribution to the WL shear angular power spectrum.}
        \label{fig4}
\end{figure}




\begin{figure}
  \centering
  \includegraphics[width=\columnwidth]{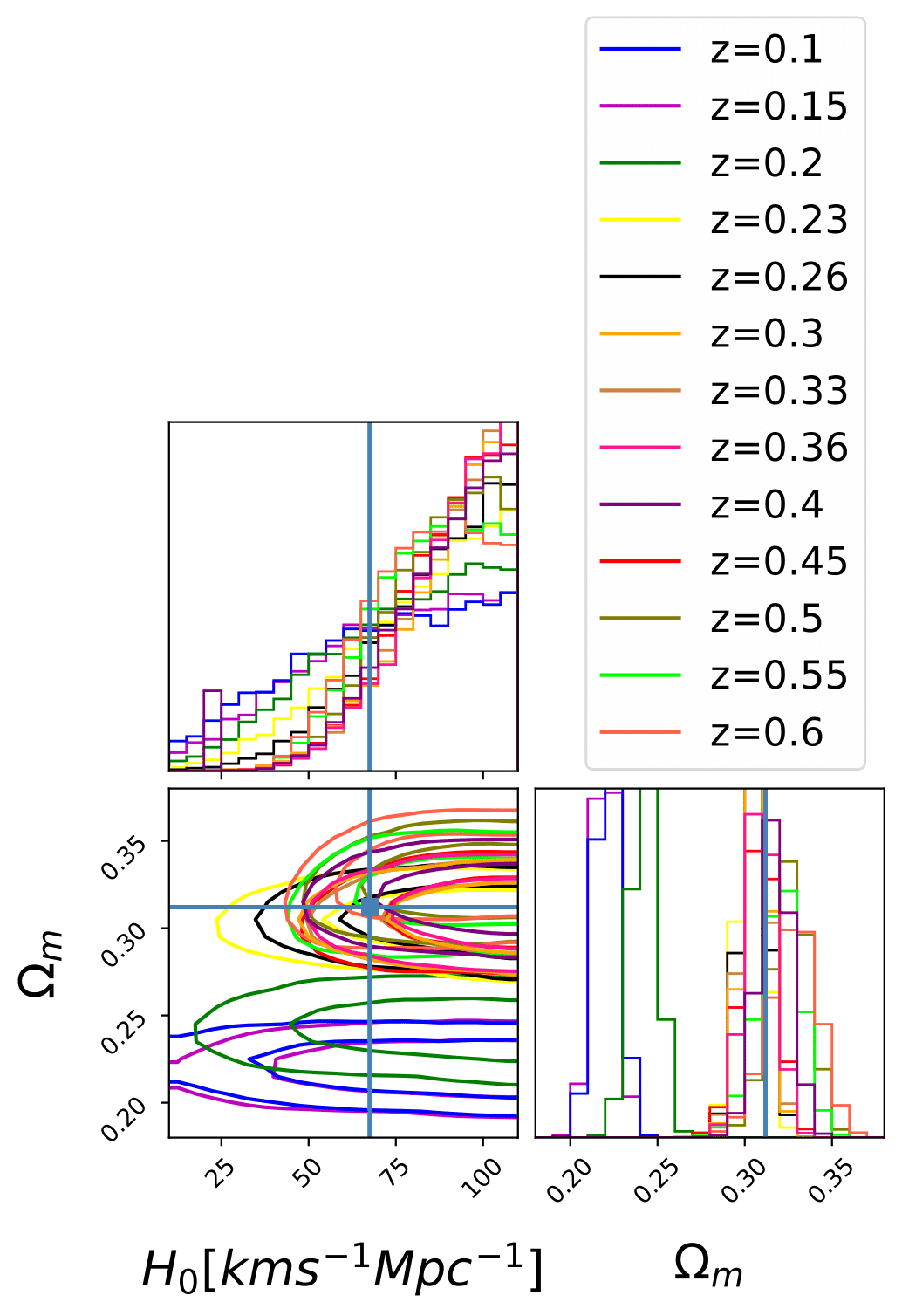}
  \caption{\justifying The expected constraints on cosmological parameters depend on the redshift of the data used. The constraints are computed using simulations with $H_0 = 67.556 \, {\rm km} \, {\rm s}^{-1}\, {\rm Mpc}^{-1}$, and $\Omega_{\rm{m}} = 0.312$. Next, a shear angular power spectrum that is theoretically derived is compared to this mock angular power spectrum based on $\Omega_{\rm{m}}$ and $H_0$.}
   \label{fig5}
\end{figure}




\begin{figure}
  \centering
  \noindent
  \includegraphics[width=\columnwidth]{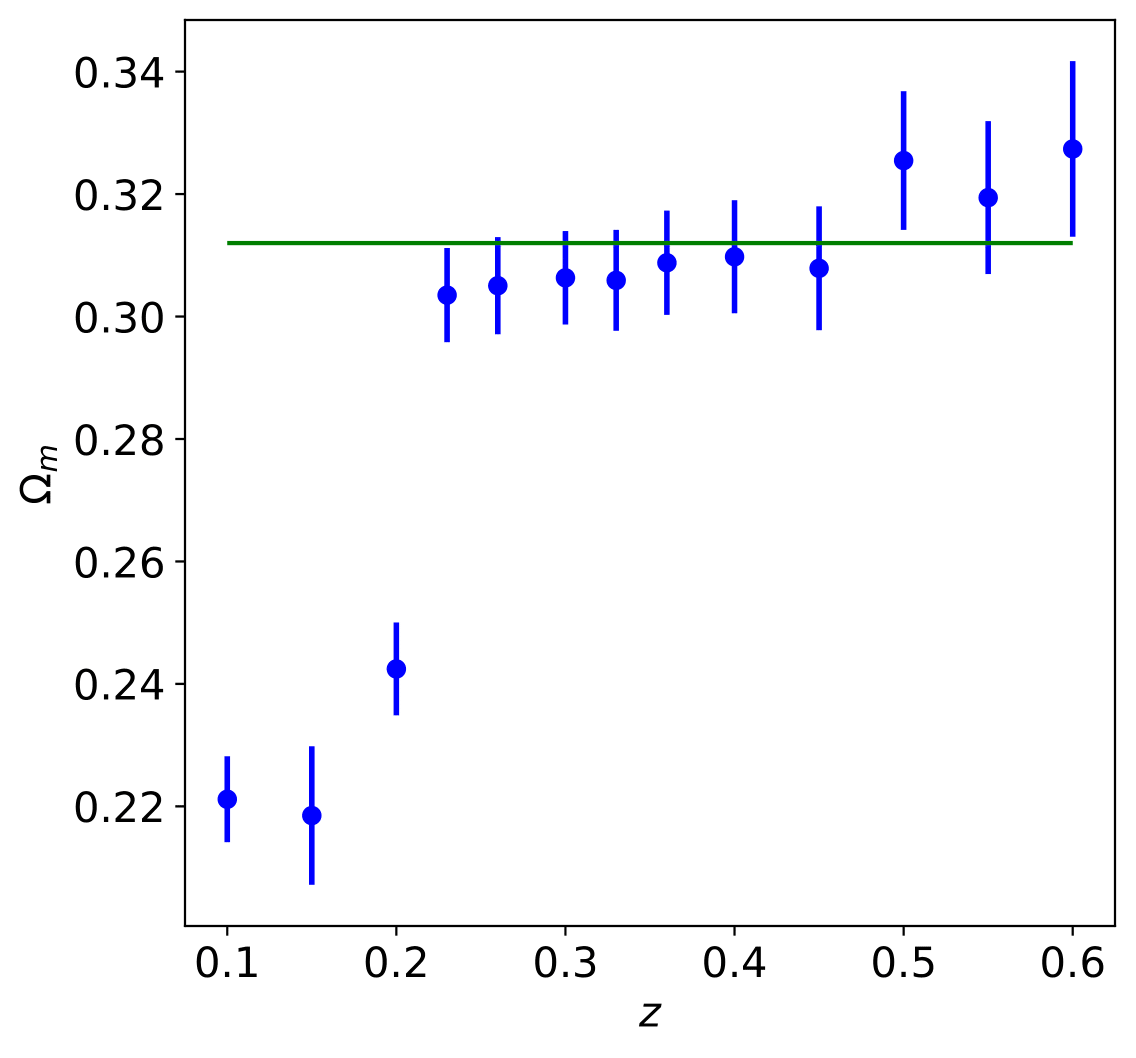}
  \caption{\justifying Constraints on the cosmological parameter $\Omega_{\rm{m}}$ with respect to source redshift. The constraints are based on the angular power spectrum of WL shear of 49,152 sources, simulated using $\Omega_{\rm{m}} = 0.312$. The constraints are analogous to those shown in Figure \ref{fig5}, but they are marginalised over $H_0$.} 
   \label{fig6}
\end{figure}



\begin{figure}
  \centering
  \noindent  \includegraphics[width=\columnwidth]{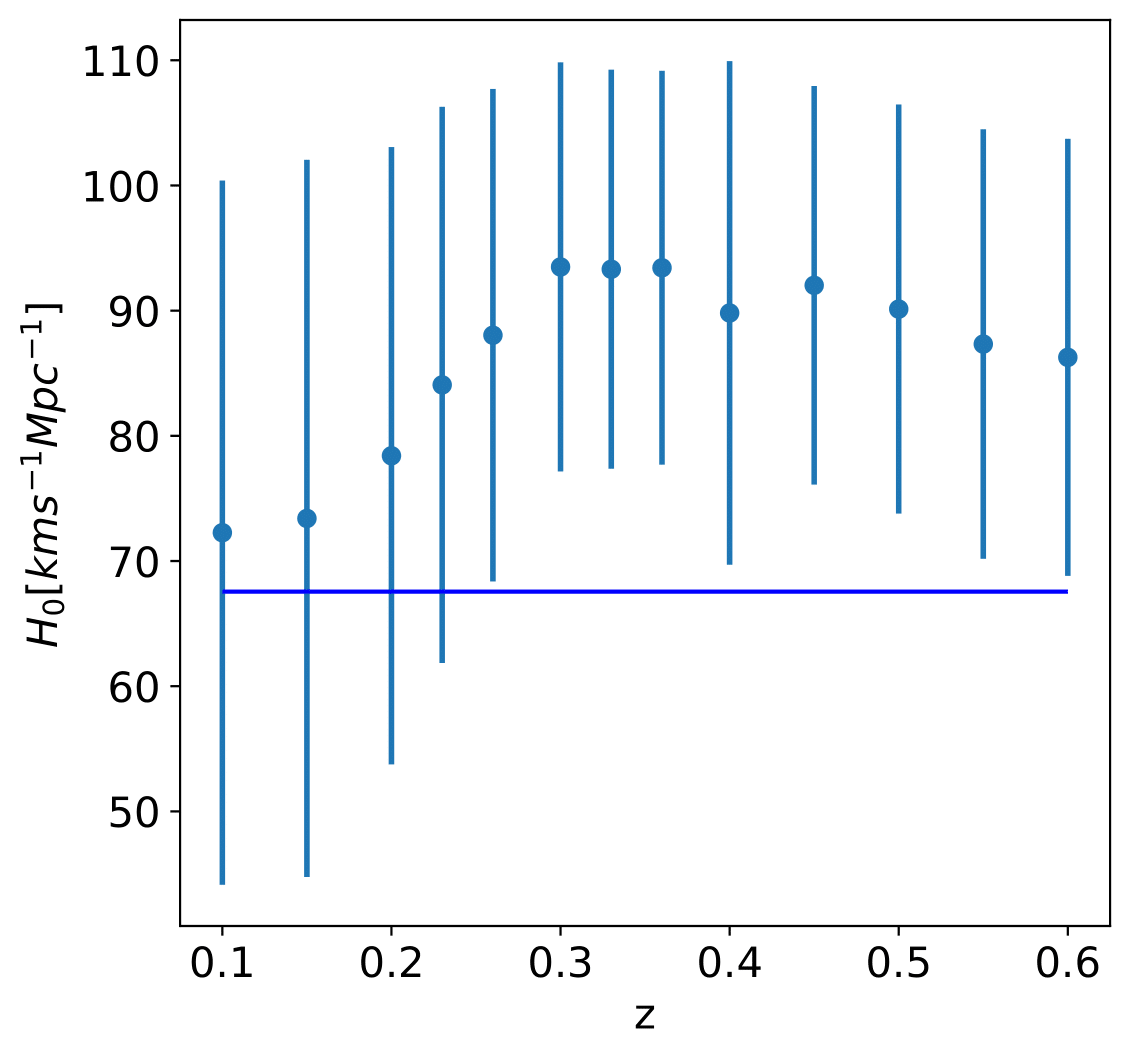}
  \caption{\justifying Constraints on the parameter $H_0$ as a function of source redshift. The constraints are based on the angular power spectrum of WL shear of 49,152 sources, simulated using $H_0 = 67.556 \, {\rm km} \, {\rm s}^{-1}\, {\rm Mpc}^{-1}$. The constraints are similar to those presented in Fig. \ref{fig5} but marginalised over $\Omega_{\rm{m}}$.}
   \label{fig7}
\end{figure}




\begin{figure}
  \centering
  \noindent
  \includegraphics[width=\columnwidth]{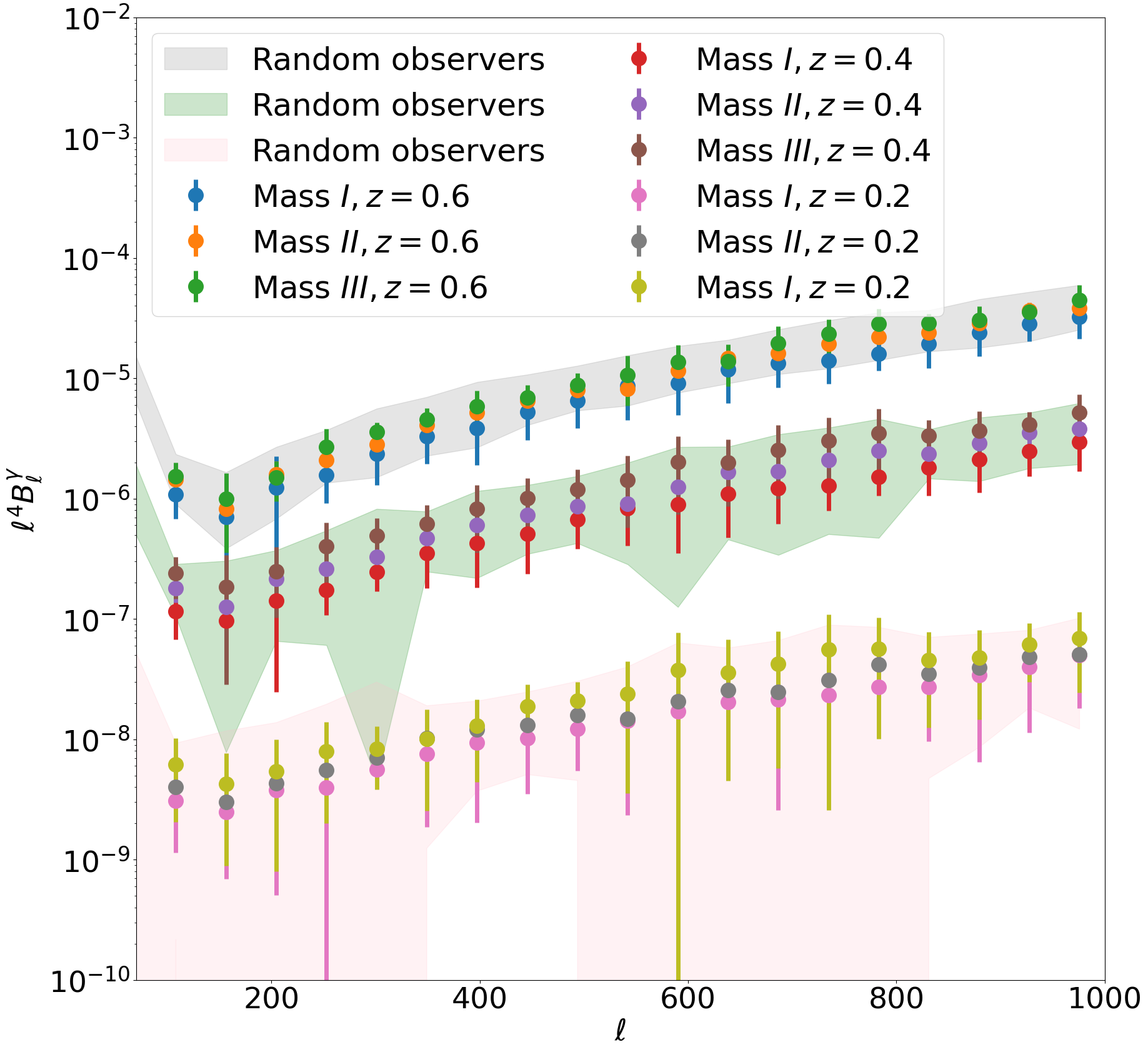}
  \caption{\justifying An analysis of the $isosceles$ configuration of the shear bispectrum with respect to redshift when observers are positioned in haloes with varying masses. In this context, markers are used to represent mean values, while error bars of varying colours are used to display the data around the mean values for separate redshifts within each mass range. The different shaded regions indicate the expected deviation band if the observer is randomly located within the simulation box.}
   \label{fig8}
\end{figure}


\begin{figure}
  \centering
  \noindent
  \includegraphics[width=\columnwidth]{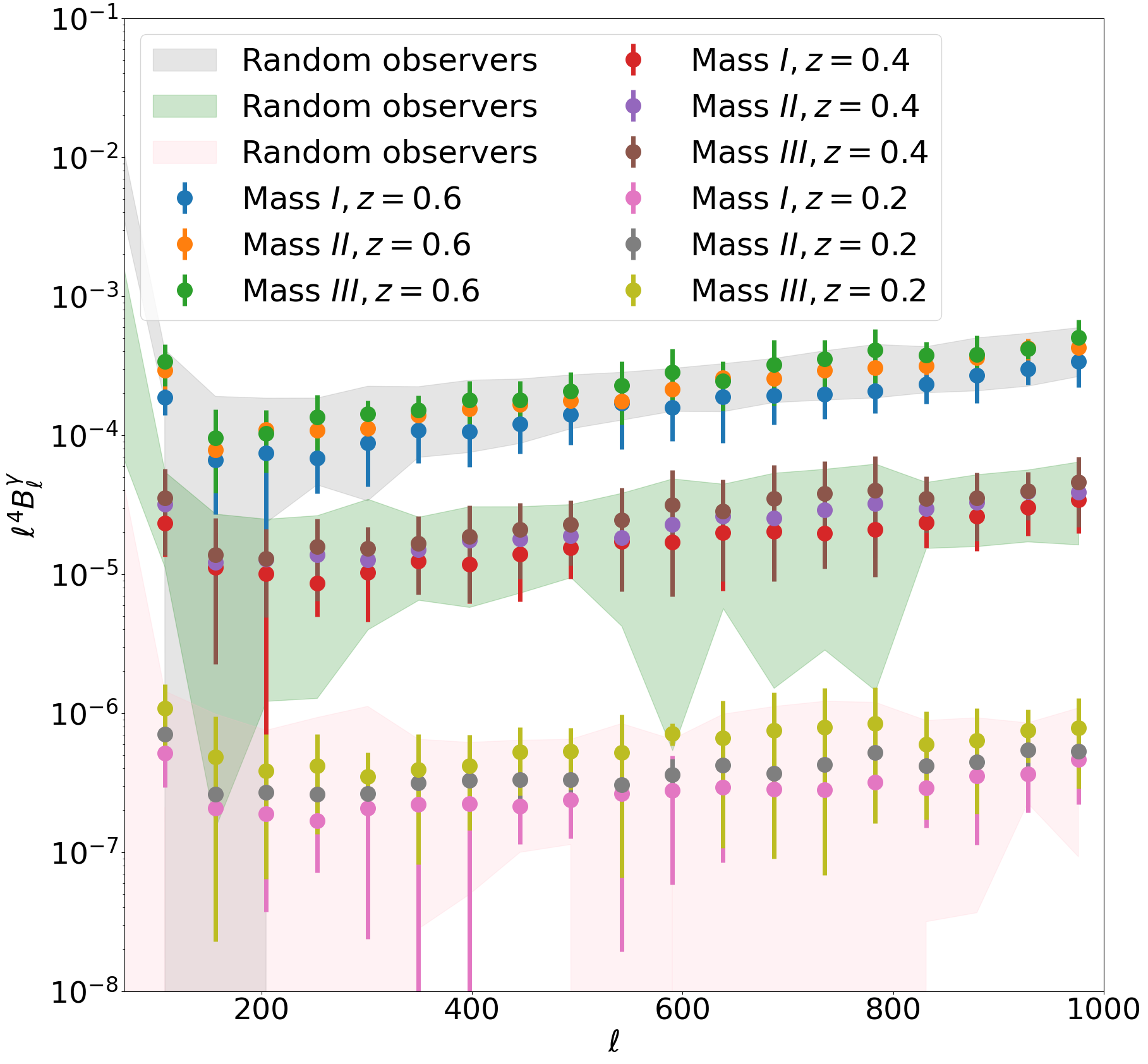}
  \caption{\justifying An analysis of the $squeezed$ configuration of the shear bispectrum with respect to redshift when observers are positioned in haloes with varying masses. In this context, markers are used to represent mean values, while error bars of varying colours are used to display the data around the mean values for separate redshifts within each mass range. The different shaded regions indicate the expected deviation band if the observer is randomly located within the simulation box.}
  \label{fig9}
\end{figure}


\begin{figure}
  \centering
  \noindent
  \includegraphics[width=\columnwidth]{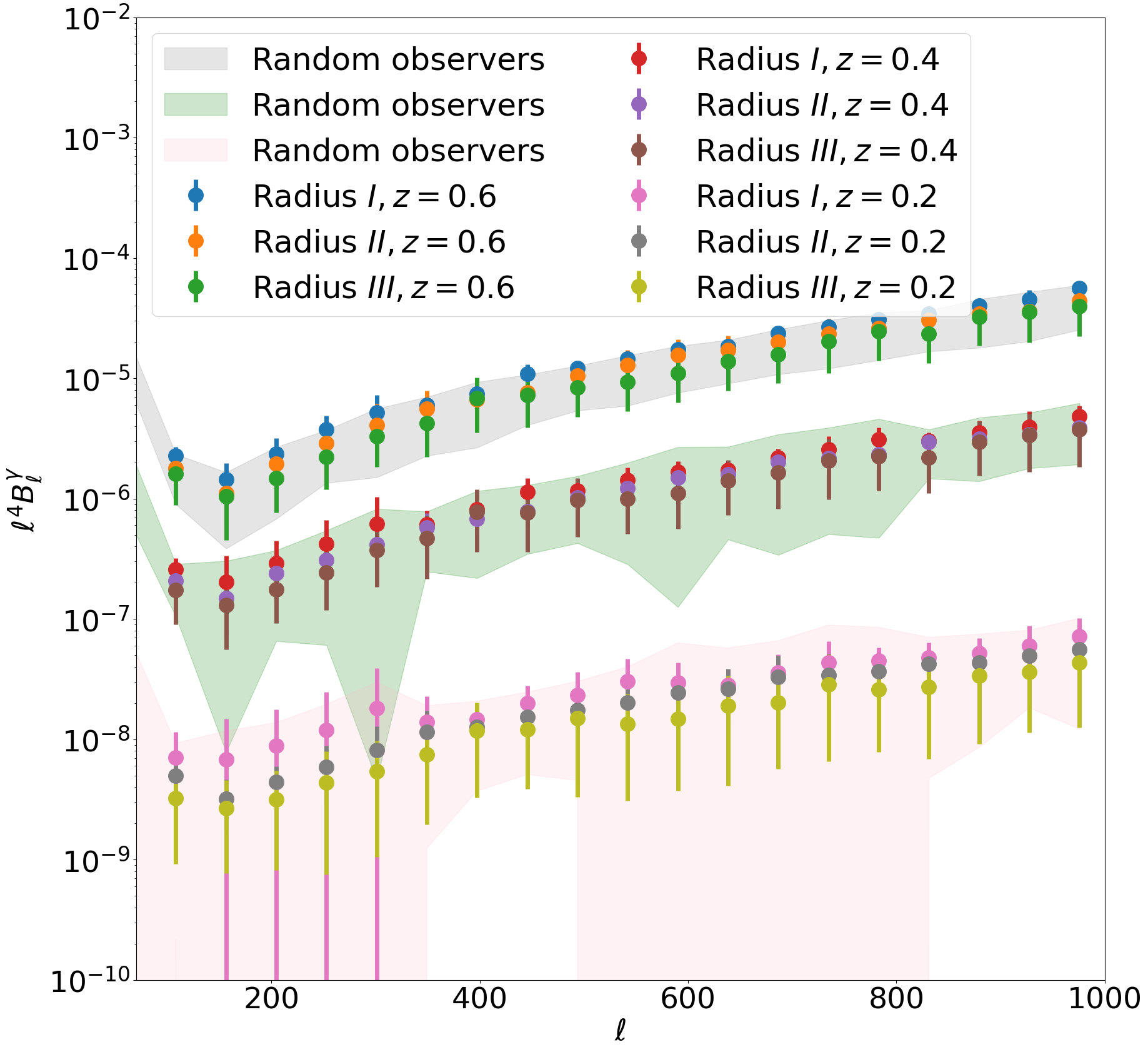}
  \caption{\justifying An analysis of the $isosceles$ configuration of the shear bispectrum with respect to redshift when observers are positioned in voids having different radii. In this context, markers are used to represent mean values, while error bars of varying colours are used to display the data around the mean values for separate redshifts within each radius range. The different shaded regions indicate the expected deviation band if the observer is randomly located within the simulation box.}
   \label{fig10}
\end{figure}


\begin{figure}
  \centering
  \noindent
  \includegraphics[width=\columnwidth]{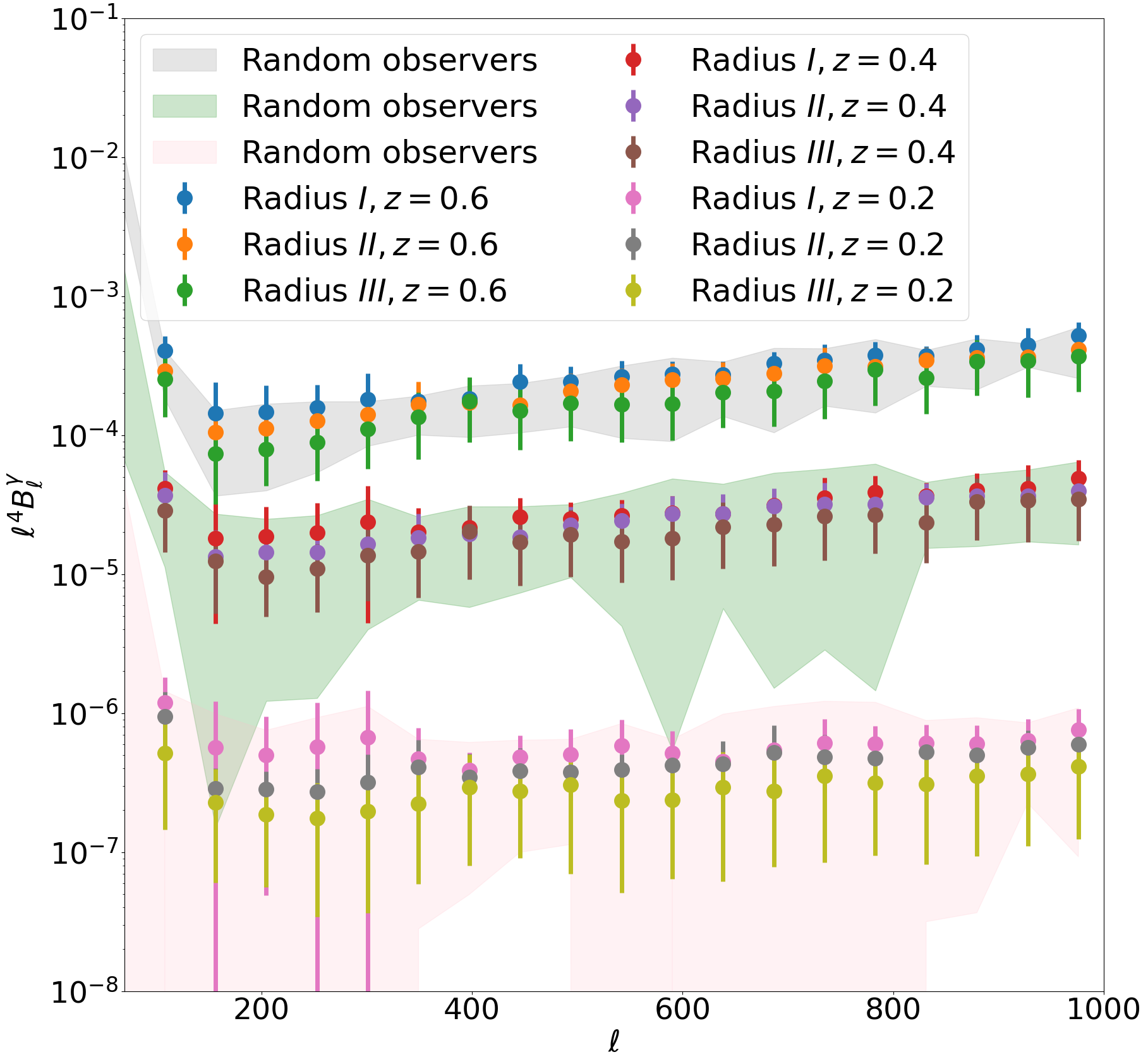}
  \caption{\justifying An analysis of the $squeezed$ configuration of the shear bispectrum with respect to redshift when observers are positioned in voids having different radii. In this context, markers are used to represent mean values, while error bars of varying colours are used to display the data around the mean values for separate redshifts within each radius range. The different shaded regions indicate the expected deviation band if the observer is randomly located within the simulation box.}
   \label{fig11}
\end{figure}



\section{WL Shear Angular power spectrum and bispectrum}
\label{sec:WL Shear Angular power spectrum and bispectrum}


\subsection{Analytical prediction of WL shear angular power spectrum}
\label{Analytical prediction of WL shear angular power spectrum}

Disregarding the sky's curvature, the Limber approximation can be taken into account to compute the WL angular power spectrum analytically for a full-sky approximation \citep{1953ApJ...117..134L, 2010A&A...523A..28K, 2017SchpJ..1232440B, 2012MNRAS.420..155K, 2017JCAP...05..014L}. Using Born approximation, the convergence can be expressed in terms of the weighted integral of the overdensity $\delta$ along the line of sight as

\begin{equation}
\begin{aligned} 
\kappa(\theta, \chi) = \frac{3 H_0^2 \Omega_m}{2 c^2} \int_0^\chi d\chi^{'}   \frac{r (\chi^{'} - \chi) r(\chi^{'})}{r(\chi)}  \frac{\delta(r(\chi^{'}) \theta, \chi^{'})}{a(\chi^{'})},
\end{aligned} 
\label{convrg}
\end{equation}

\noindent in the above equation, $\Omega_\mathrm{{m}}$ is the matter density, $\theta$ is the observed angular position, $\chi$ is the comoving distance, $H_0$ is the Hubble constant, $r(\chi)$ is the comoving angular diameter distance, and $a(\chi^{'})$ is the scale factor at $\chi^{'}$. 

In the flat-sky approximation, the convergence angular power spectrum can be expressed as 

\begin{equation}
  \langle\tilde{\kappa}(\bm{\ell}) \tilde{\kappa}^{*}(\bm{\ell}')\rangle=(2\pi)^2 \, \delta_{\rm D}(\bm{\ell}-\bm{\ell}') \, C^{\kappa}(\ell),
  \label{Fourier}
\end{equation}

\noindent Equation (\ref{Fourier}) is known as the two-point correlation function in Fourier space. Here, $C^{\kappa} (\ell)$ is the angular power spectrum of convergence and $\delta_{\rm D}(\bm{\ell})$ is the Dirac delta function. 

Since convergence can be measured from the projection of density contrast, in the Limber approximation the final equation of the convergence angular power spectra is  

\begin{equation}
  C^{\kappa}(\ell) = \int^{\chi_{\rm H}}_{0} {\rm d}\chi \frac{W(\chi)^{2}}{r(\chi)^{2}} P_{\delta}\left(k=\frac{\ell}{r(\chi)}, \chi\right),
\end{equation}
\noindent where $W(\chi)$ is the weighting function, $P_{\delta}(k,\chi)$ is the 3D matter power spectrum at the given comoving distance $\chi$,  and $\chi_{\rm H}$ is the comoving horizon distance. Again, the weighting function $W(\chi)$ can be written in the following form

\begin{equation}
  W(\chi) = \frac{3H_{0}^{2}\Omega_{{\rm m}}}{2c^2}\frac{r(\chi_{\rm H}-\chi)r(\chi)}{r(\chi_{\rm H})} \frac{1}{a(\chi)}.
\end{equation}

The two components of shear $\gamma_1$ \& $\gamma_2$ are closely related to the convergence in a flat-sky approximation, and the following equation is valid in the Fourier space \citep{2002MNRAS.335..909B}

\begin{equation}
    \tilde\gamma_1^{2} (\ell) + \tilde\gamma_2^{2} (\ell) = \tilde\kappa^{2} (\ell).
\end{equation}

\noindent Again, in the WL system, Eq. (\ref{mu1}) minimizes to 

\begin{equation}
    \mu = 1 + \delta \mu \simeq 1 + 2 \kappa,
\end{equation}

\noindent so, in the WL limit the relation between the angular power spectrum of convergence and angular power spectrum of shear can be expressed as \citep{2003MNRAS.344..789B} 

\begin{equation}
    C_{\ell}^{\delta \mu} \simeq 4 C_{\ell}^{\kappa} = 4 C_{\ell}^{\gamma},
\end{equation}

\noindent where $C_{\ell}^{\delta \mu}$ is the magnification fluctuation power, and $C_{\ell}^{\gamma}$ is the angular power spectrum of WL shear. 

Here, we measure the theoretical predictions of the WL angular power spectrum using the publicly available code \texttt{jax-cosmo}\footnote{\url{https://github.com/DifferentiableUniverseInitiative/jax_cosmo}}. We consider the results of WL angular power spectrum of shear for the multipole $\ell > 10$, because the Limber approximation is valid for small angle approximations \citep{2017JCAP...05..014L}. 

Measured shear correlations contain a shot-noise contribution due to the intrinsic ellipticities of source galaxies. Assuming that the ellipticity distribution is uncorrelated between different galaxies, the observed power spectrum between redshift bins $i$ and $j$ can be written as \citep{Kaiser1992}:
\begin{equation}
C^{\mathrm{obs}}_{ij}(\ell) = C_{ij}(\ell) + \delta_{ij}\, \frac{\sigma_\epsilon^2}{\bar{n}_i},
\end{equation}
where $\bar{n}_i$ is the average angular number density of galaxies in redshift bin $i$. The Kronecker delta $\delta_{ij}$ ensures that cross-power spectra with $i \neq j$ are not affected by shot noise. 
Thus, the cross-spectrum provides an unbiased estimator of the cosmological signal. In this expression, we have ignored additional contaminants such as observational systematics and intrinsic alignments of galaxy ellipticities. 
In our analysis, the shot-noise contribution is computed assuming a single effective source redshift bin, parametrizing the lensing volume by the sky coverage $f_{sky} = 1.6 \%$ and the rms intrinsic ellipticity $\sigma_\epsilon = 0.3$. Since we use discrete tracers such as halos and voids instead of galaxies, we replace \(\bar{n}_i\) by the tracer number density \(n_{\mathrm{tracer}}\) in Eq. (20). Here, \(n_{\mathrm{tracer}}\) is calculated as the total number of tracers in the corresponding redshifts divided by the survey solid angle \(\Omega_{\mathrm{survey}}\) (in steradians), i.e., $n_{\mathrm{tracer}} = \frac{N_{\mathrm{tracer}}}{\Omega_{\mathrm{survey}}}$. For redshift $z=0.2$, we found that $n_{\mathrm{tracer}}  = 1.5 \times 10^6$ per steradians.

\subsection{WL shear angular power spectrum from numerical simulation}
\label{WL shear angular power spectrum from numerical simulation}

The technique of measuring the angular power spectrum through the spherical harmonic coefficient can be expressed by the following equation:

\begin{equation}
\hat{C}_\ell = \frac{1}{2\ell + 1} \sum_m |\hat{a}_{\ell m}|^2,
\label{power_spectra}
\end{equation}

\noindent where, $\hat{C}_\ell$ is the angular power spectrum and $a_{\ell m}$ is the spherical harmonic coefficients.

\noindent In this study, the necessary steps to calculate the WL shear angular power spectrum are:

\begin{itemize}
    \item \texttt{compute WL shear:} we first use the Euler-Rodrigues formula to align the ray bundles initially and compute WL shear for each bundle using Eq. (\ref{shear_RBM}). 
    
    \item \texttt{WL shear maps:} we then use \texttt{healpy} \citep{2005ApJ...622..759G} to analyse the WL shear maps. We generate WL shear maps from the numerical simulations by following the \texttt{3D-RBT} algorithm \citep{2022MNRAS.509.3004E, 2022MNRAS.509.5142H, 2022MNRAS.513.5575H}. The resolution of our full-sky map is $N_{\mathrm{side}}$ = 512 which corresponds to an angular resolution of 0.11 deg. The total number of pixels that covers the entire sky is 12 $\times$ $N_{\mathrm{side}}^2$ and each map is generated in such a way so that each ray bundle get separate pixel directions on the \texttt{HEALPix} sphere \citep{1999astro.ph..5275G}.

    \item \texttt{masking effect:} the WL shear maps that we obtain from our ray tracing simulations cover about 1.6\% of the sky. \texttt{healpy} sets all masked pixels to zero, which could produce a substantial bias in the calculation of the angular power spectrum. To compute the full-sky masked angular power spectrum, we must carefully tackle this issue.

    \item \texttt{pseudo angular power spectrum:} to correct the masking effects in our analysis, we use the public code \texttt{NaMaster}\footnote{{\url{https://https://github.com/LSSTDESC/NaMaster}}}. Using this code, we obtain a pseudo angular power spectrum from the masked WL shear maps. Cosmic variance errors can be reduced by setting the band powers properly, more details regarding the code \texttt{NaMaster} can be found in this work \citep{2019MNRAS.484.4127A}.  

\end{itemize}


\begin{figure}
  \centering
  \noindent
  \includegraphics[width=\columnwidth]{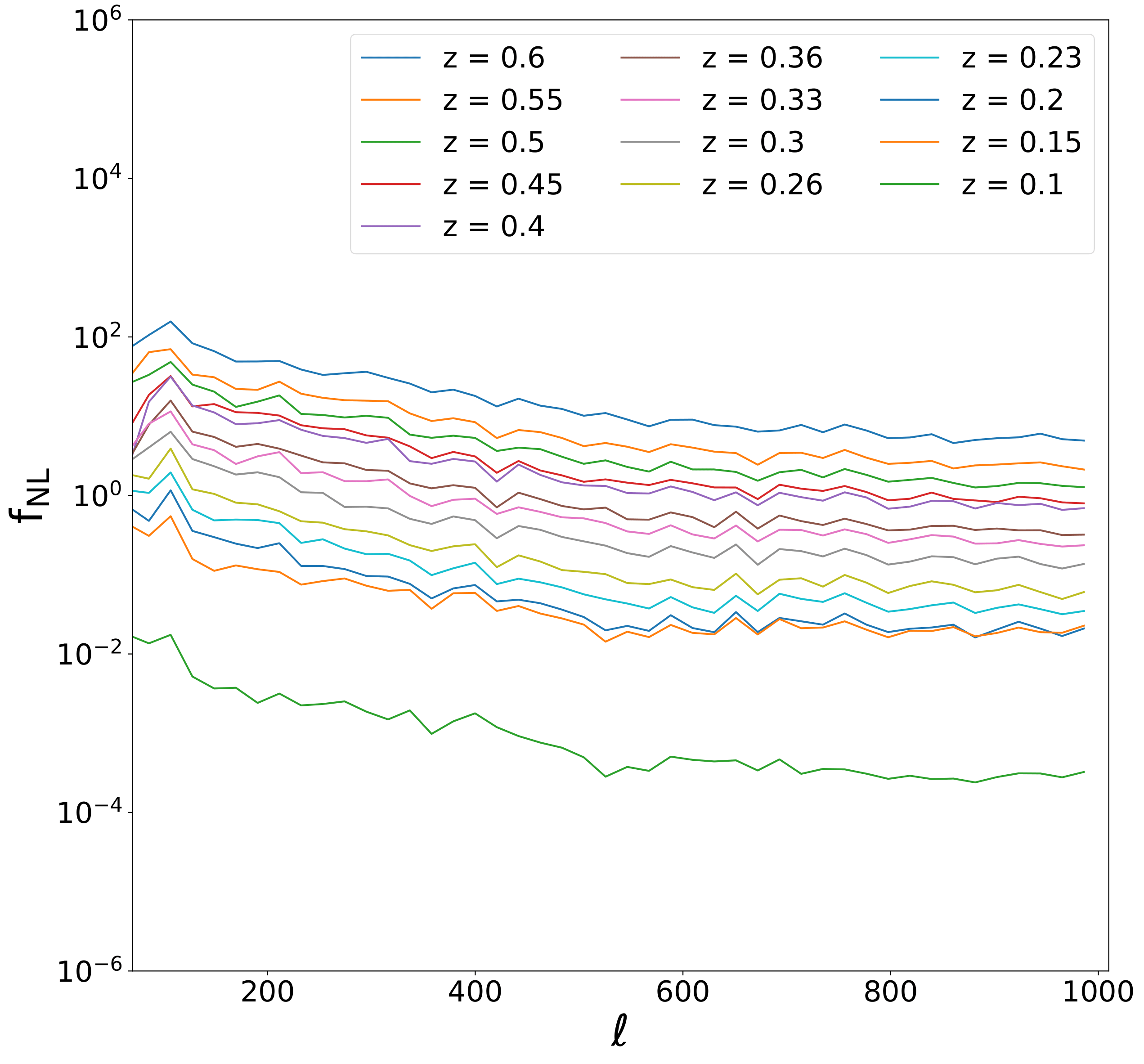}
  \caption{\justifying Non-Gaussian parameter, $f_{\rm NL}$, as a function of redshift when the observer resided within a halo of mass of approximately $10^{12.5} \rm {M}_\odot $h$^{-1}$.} 
   \label{f_nl1}
\end{figure}


To establish quantitative accuracy in the absence of a standard numerical ``random observer" baseline, we verified the fidelity of our simulation pipeline by directly comparing our results to the theoretical prediction. We calculated the standard theoretical shear PS using $\texttt{jax\text{-}cosmo}$ (Halofit) for our input cosmology. Since the Halo and Void environments induce a physical bias relative to the average universe, we cannot simply equate the measured $C_{\ell}^{\text{Halo/Void}}$ with $C_{\ell}^{\text{theory}}$. Instead, we performed the accuracy check by confirming that the average deviation induced by the environments is consistent with the statistical variance of the simulation. The simulation's quantitative accuracy is best represented by the percentage difference between the theoretical prediction and the geometric average of the halo and void results. At $\ell=1000$, we found this deviation to be $\mathbf{50\%}$. The large deviation results from using the mean of the two extreme environments (halo and void) as a proxy for the cosmic average. Because our simulation is a limited-volume realization ($f_{sky} \approx 1.6\%$), the true mean signal of this realization is expected to deviate significantly from the theoretical cosmic average due to cosmic variance. This $50\%$ accuracy, therefore, primarily reflects the cosmic variance of our small simulated volume, not a numerical error in the ray-tracing pipeline. The formula we utilized to claim accuracy at $\ell=1000$ is:$$\text{Accuracy} (\%) = \left| \frac{\left( \frac{C_{\ell}^{\text{Halo}} + C_{\ell}^{\text{Void}}}{2} \right) - C_{\ell}^{\text{theory}}}{C_{\ell}^{\text{theory}}} \right| \times 100$$

It is worth noting that for the WL shear PDFs, angular power spectra, and bispectra analyses, we computed the expected deviation band if the observer is randomly located within the simulation box. This band incorporates the $68\%$ statistical confidence intervals of the halo-centered observers and the void-centered observers, representing the maximum expected scatter of the measurement within this simulation realization.

\subsection{Analytical prediction of WL shear angular bispectrum}
\label{Analytical prediction of WL shear angular bispectrum}

The non-linearity in the case of structure formation violates the homogeneity of the growth equation and generates non-Gaussian characteristics. It is possible to study both non-linearity and non-Gaussianity from the WL statistics. Several studies focus on the modelling and computation of WL bispectrum using WL convergence and shear maps \citep{2011MNRAS.411.2241M, 2020MNRAS.493.3985M,2012MNRAS.421..797S,2014MNRAS.441.2725F}. 

The WL angular bispectrum can be expressed by the following equation which depends on the spherical harmonic coefficients of relevant fields 

\begin{equation}
\resizebox{\columnwidth}{!}{$
 \begin{aligned}
B_{\ell_1 \ell_2 \ell_3}^{\Gamma \Gamma' \Gamma''} (r_1, r_2, r_3)=\sum_{m_1,m_2,m_3} 
\langle _s{\Gamma}_{\ell_1 m_1} (r_1) _{s'}{\Gamma'}_{\ell_2 m_2} (r_2) 
 _{s''}{\Gamma''}_{\ell_3 m_3} (r_3) \rangle
\begin{pmatrix}
\ell _1 &\ell _2 &\ell _3 \\ m_1 & m_2 & m_3 \\
\end{pmatrix}.
 \end{aligned}
$}
\label{Bangleav}
\end{equation}

\noindent Where $\Gamma(r_1)$, $\Gamma'(r_2)$, and $\Gamma''(r_3)$ are fields on the surface of a sphere having spins $s$, $s'$, and $s''$ respectively. We consider full-sky approximation and add an arbitrary mask in the foreground sky. Higher order shear statistics have two modes, $electric$ E-modes and $magnetic$ B-modes, the above-mentioned bispectrum Eq. (\ref{Bangleav}) is general when the smaller contributions from the $magnetic$ B-modes will be ignored, and for full-sky approximation \citep{2011MNRAS.411.2241M} arbitrary masking could be applied. The term within the angular brackets represents the rotational invariance of the three-point correlation functions in the harmonic domain. The matrices in Eq. (\ref{Bangleav}) is the Wigner 3j-symbol, $\ell_i (i = 1,2,3)$ is the degree and $m_i(i=1,2,3)$ is the order of spherical harmonics. The Wigner 3j-symbol gives a non-zero value only for two particular triangularity conditions: i) when $|\ell_1 - \ell_2| \le \ell_3 \le (\ell_1 + \ell_2) $; ii) when $\ell_1 + \ell_2 + \ell_3$ is even.

In this study, we consider the reduced bispectrum (a normalised form of the bispectrum) \citep{2005ApJ...634...14K, 2004JCAP...08..009B} because it's commodious to use, hence 

\begin{equation}
    b_{\ell_1 \ell_2 \ell_3} = h^{-1}_{\ell_1 \ell_2 \ell_3} B_{\ell_1 \ell_2 \ell_3}, 
\end{equation}

\noindent and 

\begin{equation}
    h^2_{\ell_1 \ell_2 \ell_3} = \frac{\sum_{\ell_1} \sum_{\ell_2} \sum_{\ell_3}}{4 \pi} \begin{pmatrix} 
\ell_1 &\ell_2 &\ell_3\cr 0 & 0 & 0\cr \end{pmatrix}^2; 
\sum_{\ell_i} = (2 \ell_i +1),
\end{equation}

\noindent where $h_{\ell_1 \ell_2 \ell_3}$ is a geometric factor that doesn't depend on the WL shear maps but only depends on $\ell_i$.

We use the publicly available code \texttt{cmblensplus}\footnote{\url{https://https://github.com/toshiyan/cmblensplus}}\citep{2019PhRvD..99f3511N} to obtain the WL reduced shear bispectrum. This package is developed originally to study the non-Gaussianities in CMB lensing, but also useful in analysing WL bispectrum \citep{2020MNRAS.493.3985M}.


\begin{figure}
  \centering
  \noindent
  \includegraphics[width=\columnwidth]{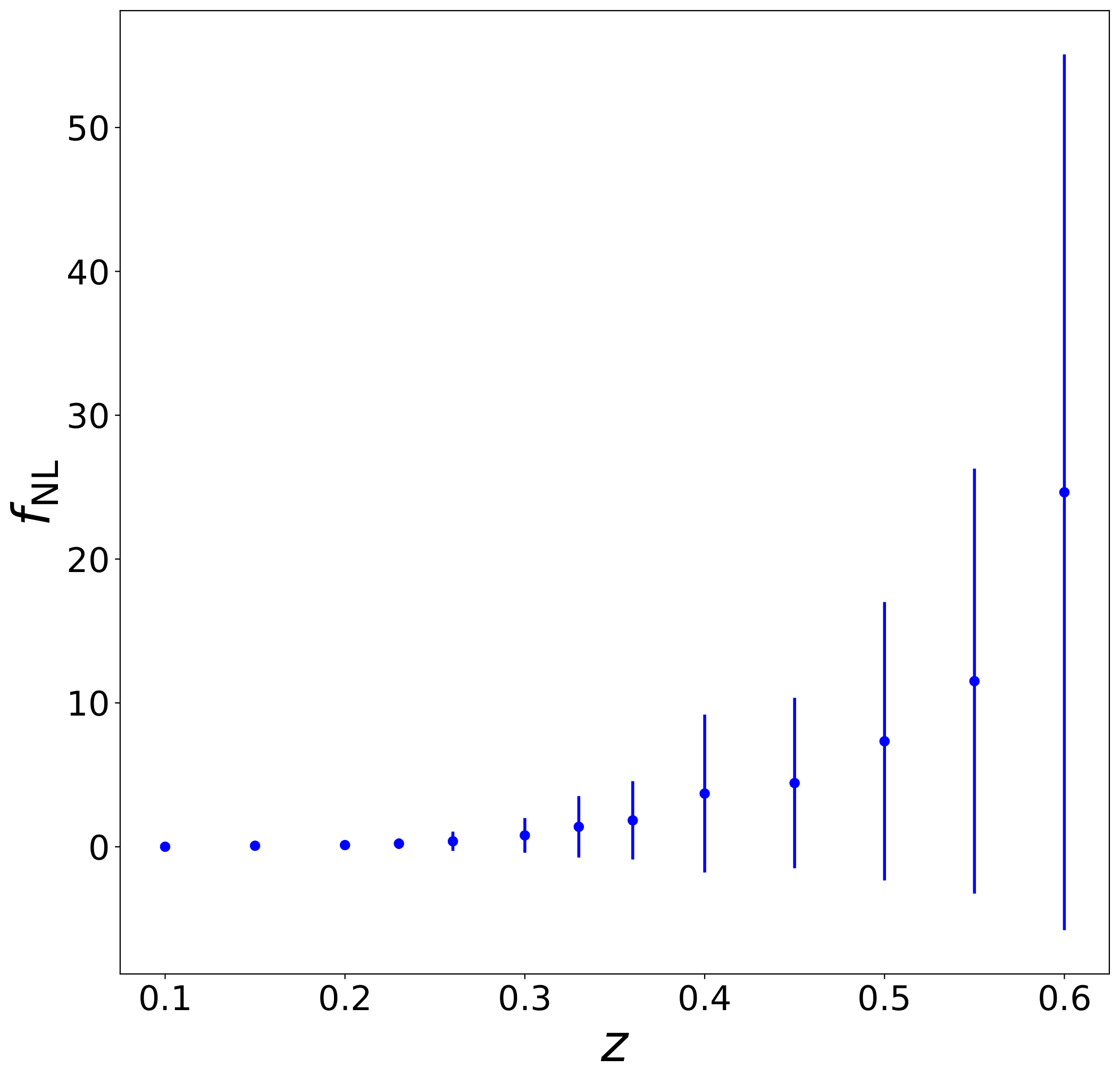}
  \caption{\justifying The inferred value of the parameter $f_{\rm NL}$.  The intervals represent 68\% spread of possible values inferred by random observers placed across the simulation but reside within a halo of mass of approximately $10^{12.5} \rm {M}_\odot h^{-1}$.}
   \label{f_NL2}
\end{figure}


%
%


\subsection{WL shear angular bispectrum from numerical simulations}
\label{WL shear angular bispectrum from numerical simulations}

We infer the WL shear angular bispectrum from simulation data using the public code \texttt{cmblensplus}. First, we measure the angular power spectrum from the arbitrarily masked map of WL shear and then measure the WL angular bispectrum from the masked shear angular power spectrum data. For full-sky approximation, the unbinned reduced bispectrum estimator here follows the following expression 

\begin{equation}
 \begin{aligned}
b_{\ell_1\ell_2\ell_3} = h^{-2}_{\ell_1 \ell_2 \ell_3} \bigg \langle \int d^2 \hat{\mathbf{n}} \kappa_1 (\hat{\mathbf{n}})  \kappa_2 (\hat{\mathbf{n}})  \kappa_3 (\hat{\mathbf{n}})
\bigg \rangle,
 \end{aligned}
 \label{unbinned_bispectrum}
\end{equation}

\noindent where $\kappa_i(\hat{\mathbf{n}})$ can be obtained from the inverse harmonic transformation of relevant fields. 



\noindent We compute binned reduced bispectrum here to minimize the simulation errors from cosmic variance. We refer this paper \citep{2019PhRvD..99f3511N} to know more about the derivation of reduced binned bispectrum. There are four specific configurations of WL bispectrum:

\begin{itemize} 
\item $equilateral$ configuration : $\ell_1 = \ell_2 = \ell_3 = \ell$
\item $folded$ configuration : $\ell_1 = 2\ell_2 = 2\ell_3 = \ell$
\item $isosceles$ configuration : $\ell_1 = \ell, \ell_2 = \ell_3 = 500$
\item $squeezed$ configuration : $\ell_1 = 100, \ell_2 = \ell_3 = \ell,$
\end{itemize} 

\noindent whereas, we focus on two specific configurations ($isosceles$ \& $squeezed$) in this study because these two configurations are mostly sensitive to the evolution history of the LSS. 

\section{Results}
\label{sec:Results}

\subsection{Probability distribution functions and the local environment}
\label{Probability distribution functions and the local environment}

We use the particle's snapshot from \texttt{gevolution} at $\it z = 0 $ to identify the cosmic structures. We use the algorithm \texttt{ROCKSTAR} to identify haloes and another algorithm \texttt{Pylians} for cosmic void identification. Then the haloes and voids are classified according to their masses and radii. An Observer's position is an important thing while analysing the influence of the local environment and we put the observers here in different haloes and voids and analyse how their observations vary depending on their positions on the LSS. We calculate PDFs by solving 49152 bundles of geodesics using the $\texttt{3D-RBT}$ method. For each halo mass range and void radii range, we then evaluate the mean PDFs and error bars from the 68\% data around mean PDFs.

We show the PDFs of cosmic shear at different redshifts for different halo mass ranges in Figure \ref{fig1}. Figure \ref{fig2} reflects the result of how the WL shear changes with respect to redshifts when observers are positioned in void regions having varying radii. In both figures, the error bars show 68\% data around mean values and the gray-shaded regions indicate the deviation band if the observer is randomly located within the simulation box. With the increasing of redshifts, the peaks of the PDFs become higher but the shapes do not change a lot (\citep{2011ApJ...742...15T} noted the same outcome). Our study suggests that the observers in massive haloes will observe higher cosmic shear as compared to the observers residing in less massive halo regions (see Figure \ref{fig1}). Again, the analysis of PDFs of voids having different radii suggests that smaller voids may experience higher cosmic shear as compared to the bigger voids (see Figure \ref{fig2}). These properties can be explained by the concept of particles' density distributions within cosmic structures. Massive halo regions are commonly high in density in comparison to less massive haloes. It is believed that voids are formed in the early universe due to the quantum mechanical fluctuations in the density distributions; these voids aren't empty but regions with low density. Generally, larger voids possess lower particle densities compared to smaller voids. Because larger voids form where matter is not compressing into dense structures. Consequently, their average particle densities are lower than those of smaller voids. We find observers located in denser environments may observe more distortion of light rays than the observers located in less dense environments, i.e., for highly dense environments the amplitude of the PDFs is also higher.

\subsection{Angular power spectrum and the local environment}
\label{Angular power spectrum and the local environment}

Figure \ref{fig3} shows how the WL shear angular power spectrum varies at different redshifts when the observers reside in haloes with varying masses. We set the observers within haloes (voids) with varying masses (radii) and solved 49152 ray bundles. How the shear angular power spectrum changes at different redshifts for observers located in voids having different radii has been depicted in Figure \ref{fig4}. 
In both figures, we use the markers to specify the mean values, while error bars display the data around the mean values. We find here the same properties as we have seen from the analysis of PDFs (please see Section \ref{Probability distribution functions and the local environment}). The different shaded regions indicate the expected deviation band if the observer is randomly located
within the simulation box. Due to the density variation in different halo regions having different masses and in different void regions with varying radii, the angular power spectrum of WL shear goes down when the observer is moving from the highly dense regions towards the less dense regions. We also see that the amplitude of the angular power spectrum is higher as light travels towards higher redshifts. At higher redshifts, the measurements are more statistically significant because a larger number of distant galaxies (sources) contribute to the lensing signal. Furthermore, the shear angular power spectrum exhibits larger amplitudes for these high-redshift sources due to the increased integrated path length and lensing efficiency, even though the large-scale structures themselves are less evolved (less non-linear) at these earlier times.


\subsection{Constraints on cosmological parameters from the angular power spectrum}

Here we particularly focus on how the local environmental dependency varies with redshift or whether is there any minimum redshift after which we can neglect this effect. To achieve this goal - firstly, we randomly choose one halo having a mass of $10^{12.5} \mathrm{M}_\odot h^{-1}$ and then place an observer there. The reason for randomly picking a halo from that mass range is that the mass is close to the Milky Way. Using the ray tracing algorithm $\texttt{3D-RBT}$ method, we then compute the angular power spectrum for different source redshifts numerically from the WL cosmic shear map data (see section \ref{WL shear angular power spectrum from numerical simulation} to know more). The strategy of generating model data has been described in section \ref{Analytical prediction of WL shear angular power spectrum}. In the case of WL study, both $S_8$ and $\sigma_8$ are crucial quantities as they directly impact the magnitude of lensing signals. But in \texttt{gevolution}, the parameter, $\sigma_8$, is hard-coded, and we fix another parameter, $A_s$, in our simulation. So, we estimate the posterior distributions of two cosmological parameters, $\Omega_{\mathrm{m}}$ \& $H_0$, for a given parameter space by adopting the well-known Markov chain Monte Carlo (MCMC) technique. We implement the python package \texttt{emcee}\footnote{\url{https://github.com/dfm/emcee}} here, and the uniform prior ranges for the cosmological parameters are $\Omega_{\mathrm{m}}: [0.1, 1.2]$ \& $H_0:[30, 120]$. Figure \ref{fig5} reflects the outcome based on our likelihood analysis of cosmological parameters to see the local environment's impact on WL shear. This figure shows how the posterior distributions of $\Omega_{\mathrm{m}}$ and $H_0$ change at different source redshifts, and blue line indicate our fiducial cosmology here i.e., $H_0 = 67.556 \, {\rm km} \, {\rm s}^{-1}\, {\rm Mpc}^{-1}$ and $\Omega_{\mathrm{m}} = 0.312$. We can comment on the local environment's effect on WL shear from this 2D contour plot and in total 49152 sources are taken into account to generate all posteriors at all redshifts, though the number of sources change with redshifts for real-world surveys. 
Figure \ref{fig6} shows how the constraint on the cosmological parameter $\Omega_m$ changes at different redshifts. We find tighter constraints on the parameter $\Omega_{\mathrm{m}}$ over redshift $z = 0.2$, and before this redshift the constraints disregard the true value. This significant bias could be the reason for the local environment's impact. On the other hand, the sensitivity to the cosmological parameter $H_0$ is not visible. This is understandable as we can see that there is an ongoing inconsistency in the measured value of Hubble parameter $H_0$ by the realistic surveys (for example - $H_0 = 74.03 \pm 1.42 \, {\rm km} \, {\rm s}^{-1}\, {\rm Mpc}^{-1}$ has been measured by \texttt{SHOES} collaboration \citep{2019ApJ...876...85R}, $H_0 = 67.27 \pm 0.60 \, {\rm km} \, {\rm s}^{-1}\, {\rm Mpc}^{-1}$ has been measured by Planck collaboration \citep{2020A&A...641A...6P}, and $H_0 = 67.44 \pm 0.58 \, {\rm km} \, {\rm s}^{-1}\, {\rm Mpc}^{-1}$ has been measured by \cite{2019arXiv191000483E}). It should be noted that these findings are consistent with the results reported in \cite{2021MNRAS.505.4935H}, where it was argued that WL shear can not constrain the Hubble parameter $H_0$ because the WL shear angular power spectrum changes with the physical length scale but not with the Hubble parameter $H_0$. 

\subsection{Bispectrum and the local environment}

After identifying the positions of haloes and voids from the particles' snapshot, we set the observers within different haloes and voids and solved null geodesics. The techniques of computing WL shear bispectrum analytically as well as numerically have been explained in Sections \ref{Analytical prediction of WL shear angular bispectrum} \& \ref{WL shear angular bispectrum from numerical simulations}. Figure \ref{fig8} (Figure \ref{fig9}) shows the $isosceles$ ($squeezed$) configuration of the WL shear bispectrum for observers located in haloes having different masses.  The changes of WL shear bispectra when the observers are lying within void positions having different radii are depicted in Figure \ref{fig10} ($isosceles$ configuration) \& in Figure \ref{fig11} ($squeezed$ configuration). For all of the above-mentioned figures (Figs. 9-12) - markers indicate mean values, whereas error bars of varying colours display data indicating 68\% deviation from the mean value at each mass range or each void radius range and the different shaded regions indicate the expected deviation band if the observer is randomly located within the simulation box. We find $squeezed$ bispectra has a higher amplitude than the $isosceles$ configurations for all haloes and voids. We observe the WL shear bispectrum vary in the same way as we found from the analysis of PDFs (refer to Section \ref{Probability distribution functions and the local environment}) \& angular power spectrum (refer to Section \ref{Angular power spectrum and the local environment}). So, for every case of WL shear bispectra, we find the more dense the environment the more high the bispectrum amplitude. 



\subsection{The role of non-Gaussian parameter}
In the flat-sky limit, the estimator of a local primordial signal can be written as \citep{2005ApJ...634...14K, 2000PhRvD..62d3007H}
\begin{equation}
 \begin{aligned}
   {\cal E}_{\rm loc} = \frac1{\cal N}\int \frac{d^2 l_1}{(2 \pi)^2} \frac{d^2 l_2}{(2 \pi)^2} \frac{d^2 l_3}{(2 \pi)^2} (2 \pi)^2 \delta(\vec l_1 +\vec l_2 +\vec l_3) \\
\frac{B_{\rm loc}(l_1,l_2,l_3)}{6 C_{l_1} C_{l_2}C_{l_3}} a(\vec l_1) a(\vec l_2)a(\vec l_3) \;, \label{loc_estimator}  
 \end{aligned}    
\end{equation}
\noindent where $B_{\rm loc}(l_1,l_2,l_3)$ is the bispectrum for local non-Gaussianities when $f_{\rm NL}^{\rm loc}=1$. The unbiased condition of the estimator is $\langle {\cal E}_{\rm loc} \rangle = f_{\rm NL}^{\rm loc}$, and a normalization factor is needed to initiate to make the estimator unbiased. So, we can write
\begin{equation}
    {\cal N} = \frac{1}{\pi} \int  \frac{d^2 l_2 \, d^2 l_3}{(2 \pi)^2} \frac{[B_{\rm loc}(l_1,l_2,l_3)]^2}{6 C_{l_1} C_{l_2}C_{l_3}} \;,
\end{equation}
\noindent where $\cal N$ is the normalization parameter. If we apply the local estimator (see Eq. \ref{loc_estimator}) to an arbitrary bispectra signal $B_X (l_1,l_2,l_3)$, then $f_{\rm NL}^{\rm loc}$ can be expressed as \citep{2011JCAP...11..025C}
\begin{equation}
  f_{\rm NL}^{\rm loc} = \frac1{\cal N} \cdot \frac{1}{\pi} \int  \frac{d^2 l_2 \, d^2 l_3}{(2 \pi)^2} \frac{B_{\rm loc}(l_1,l_2,l_3) B_X (l_1,l_2,l_3)}{6 C_{l_1} C_{l_2}C_{l_3}} \;, \label{contamination}  
\end{equation}
\noindent where $f_{\rm NL}^{\rm loc}$ is known as the non-Gaussian parameter. The following equation is helpful to solve the above-mentioned equation (Eq. \ref{contamination})
\begin{equation}
 \begin{aligned}
   \hspace{1cm}\int d^2 l_2 \, d^2 l_3 \, F(l_1,l_2,l_3) = \\
   4 \pi \int \frac{ l_1 dl_1\, l_2 dl_2\,l_3 dl_3}{(-l_1^4 -l_2^4 -l_3^4 + 2 l_1^2 l_2^2 + 2 l_2^2 l_3^2 + 2 l_1^2 l_3^2)^{1/2}} 
   F(l_1,l_2,l_3)\;, \label{extra}
 \end{aligned}
\end{equation}
\noindent where $F$ is a function of the moduli of the multipoles.

Finally, we solve Eq. \ref{extra} for the $squeezed$ bispectra configuration by considering these triangular inequalities conditions: $2 \le l_1 \le 100$, $20\, l_1\le l_2 \le l_{\rm max}$ and $l_2-l_1 \le l_3 \le {\rm min}(l_1+l_2,l_{\rm max})$, and then compute the non-Gaussian parameter, $f_{\rm NL}^{\rm loc}$ from Eq. \ref{contamination}. The unbiased format of the local estimator is shown in Equation \ref{loc_estimator}. We use an arbitrary bispectra signal, $B_X (l_1,l_2,l_3)$, and then estimate the parameter $f_{\rm NL}^{\rm loc}$ from Equation \ref{contamination} to find out how contaminated the system is. We follow the procedures mentioned in Sections \ref{Analytical prediction of WL shear angular bispectrum} \& \ref{WL shear angular bispectrum from numerical simulations} to compute the WL bispectra. We choose the $squeezed$ configuration because the primordial local signal is at its peak in this region.

We repeat this procedure for every random observer located in a halo of mass $10^{12.5} \rm {M}_\odot \textit{h}^{-1}$.
How the non-Gaussian parameter, $f_{\rm NL}$ varies with redshift has been demonstrated in Figure \ref{f_nl1}. Through the calculation of $f_{\rm NL}$ as a function of angular scale, it becomes possible to examine whether the degree of non-Gaussianity differs at various angular scales. This investigation may yield valuable insights into the fundamental physical mechanisms that gave rise to the fluctuations in primordial density distributions during the early universe. Our findings presented in Figure \ref{f_NL2} suggest some small bias in the inference of the parameter $f_{\rm NL}$, which would be useful in the estimation of uncertainty on the measure of $f_{\rm NL}$ based on the WL shear bispectrum.

\section{Discussion and outlook}
\label{sec:Discussion and outlook}

In this paper, we employed the relativistic $N$-body code \texttt{gevolution} to explore the local environmental dependence on WL shear statistics. Our findings show that the cosmological parameter, $\Omega_{\mathrm{m}}$, is negligibly affected by the presence of the local environment for redshifts $z > 0.2$, suggesting that the bias arises from the matter correlations becoming insignificant at higher redshifts. Furthermore, the bispectrum analysis discloses that the average influence of this local environment on the non-Gaussian parameter, $f_{\rm NL}$, is consistent with zero. Despite that, the presumption of $f_{\rm NL}$ esteems as high as 10 could be possible from specific observers' locations within the LSS. These results highlight the necessity of considering the local environmental impacts when studying the WL shear statistics, particularly for future LSS surveys that require precision at the percentage level. In addition to offering inferences that could enhance the accuracy of upcoming cosmological analyses, this work reveals a skeleton for comprehending potential biases in cosmological parameter estimation.  

We begin from the weak gravitational potentials generated from the relativistic $N$-body code \texttt{gevolution}. We then identify the positions of the cosmic structures/local environments using public algorithms \texttt{ROCKSTAR} \& \texttt{Pylians} from the simulation volume. Finally, we place observers inside different local cosmological environments, solve null geodesic equations using the ray tracing algorithm $\texttt{3D-RBT}$, and generate the WL shear map data. We compute one-point, two-point, and three-point correlation functions from the cosmic shear map data and study the impacts of the local environment. We also run MCMC samplings using WL shear mock data from the simulation, compute constraints on cosmological parameters as a function of redshifts, and analyse the effects of our local density contrast on WL properties. 

We give an estimation of how much WL shear impacts by the local density distributions at the observers' position. Although the local environment does not influence the observations of an ideal FLRW, the perspective of a realistic observer is dependent on the cosmological environment in which they are located.
From this study, we conclude that below the redshift $z$ = 0.2, the local environment's influence on the constraint on the cosmological parameter $\Omega_{\mathrm{m}}$ is noticeable. However, the local environment's effect is less noticeable and can be neglected above this redshift, i.e., redshift $z$ = 0.2. So, it is important to consider the local environment's influence while observing data at low redshift. 
This result coincides with the conclusion about the constraint on the parameter $\Omega_{\mathrm{m}}$ from our previous study \citep{2022MNRAS.509.3004E}. But we observe different results from the constraint on the Hubble constant $H_0$. We find that WL shear provides almost no constraint on the parameter $H_0$ which means WL shear observations are insensitive to $H_0$. Nowadays, non-identical values of $H_0$ are measured by different observations, so our result is also meaningful from this point of view. Recently, this paper \citep{2021MNRAS.505.4935H} has mentioned that data from the current WL shear cannot improve the constraints on $H_0$ and he has also explained the possible reasons behind this. The rescaling of $H_0$ can change the matter power spectrum as well as the WL shear correlations. Several studies have been done already, where the authors described how the sensitivity to $H_0$ abandons by discarding all the details from the matter power spectra \citep{2003PhRvL..91n1302J, 2007MNRAS.374.1377T, 2004ApJ...600...17B, 2005ApJ...635..806Z}. We calculate all of the results reported in this paper using 49152 sources. 

To observe the LSS at a better precision is the topmost priority of all the present and upcoming WL surveys. The result of this research is a feasible investigation that emphasises the significance of accounting for the influence of the local environment when collecting data from upcoming surveys. The cosmic shear signals are non-Gaussian and by unfolding the higher-order cumulants in detail, we could get more meaningful results. It is worth noting that the statistical uncertainties in our analysis are limited by the simulation's field of view ($f_{\rm sky} \approx 1.6\%$). Since measurement errors scale approximately as $\propto 1/\sqrt{f_{\rm sky}}$, current and future wide-field surveys (e.g., DES, KiDS, Euclid) covering larger sky fractions will yield significantly tighter constraints than those presented here.
We find that in the case of $squeezed$ bispectra configuration, the influence of the local environment on the non-Gaussian parameter $f_{\rm NL}$ is minimal. We numerically calculate the constraints on cosmological parameters based on one observer's observation, though other results are from the study of multiple observers. A stronger analysis of the local environment's influence on WL statistics would consider a group of observers. It would also be interesting to analyse whether other configurations of the higher-order statistics of WL shear (e.g., bispectrum) can improve the sensitivity to the Hubble parameter $H_0$. We left this interesting question for our future study. Again, using MCMC analysis, we compute the constraints on only two cosmological parameters, $\Omega_m$ and $H_0$. Because to statistically explore the parameter space, MCMC simulations require the generation of a large number of samples, and an increase in parameters causes an exponential increase in computational cost. Constraining more parameters (e.g., $S_8$) could be an interesting addition to this work because lensing constraints could be helpful to obtain good precision, break parameter degeneracies, and gain deeper knowledge to understand the local environment's impact on WL statistics.

\section*{ACKNOWLEDGMENTS}
We thank the anonymous referee for his/her comments, which improved the quality of the manuscript. The authors would like to thank Toshiya Namikawa for their comments regarding the computation of bispectrum using \texttt{cmblensplus}. We would like to express our gratitude to Artemis, the HPC support at the University of Sydney, for providing the computational resources that have contributed to the production of the results presented in this paper. SAE is supported by the Commonwealth Government funded Research Training Program (RTP) Stipend Scholarship. This work has made use of \texttt{numpy} \citep{2020Natur.585..357H}, \texttt{h5py}\footnote{\url{https://www.h5py.org/}}, \texttt{matplotlib} \citep{2007CSE.....9...90H}, \texttt{mpi4py}\footnote{\url{https://github.com/mpi4py/mpi4py}}, and \texttt{eqtools}\footnote{\url{https://eqtools.readthedocs.io/en/latest/}}.

\section*{DATA AVAILABILITY}
The data that support the findings of this article are not publicly available. The data generated as part of this project may be shared on a reasonable request to the corresponding author.

\nocite{*}


\end{document}